\documentclass{elsart}

\usepackage{graphicx}
\usepackage{amsmath}
\usepackage{amssymb}

\journal{Annals of Physics}

\begin{document}

\begin{frontmatter}

\title{Mean field at finite temperature and symmetry breaking}
\author{A. Beraudo}, 
\author{A. De Pace}, 
\author{M. Martini} and
\author{A. Molinari}
\address{Dipartimento di Fisica Teorica dell'Universit\`a di Torino and \\ 
  Istituto Nazionale di Fisica Nucleare, Sezione di Torino, \\ 
  via P.Giuria 1, I-10125 Torino, Italy}

%\date{\today}

\begin{abstract}
For an infinite system of nucleons interacting through a central
spin-isospin schematic force we discuss how the Hartree-Fock theory at finite
temperature $T$ yields back, in the $T=0$ limit, the standard zero-temperature
Feynman theory when there is no symmetry breaking.
The attention is focused on the mechanism of cancellation of the higher order
Hartree-Fock diagrams and on the dependence of this cancellation upon the range
of the interaction. 
When a symmetry breaking takes place it turns out that more iterations are
required to reach the self-consistent Hartree-Fock solution, because the
cancellation of the Hartree-Fock diagrams of order higher than one no longer
occurs. We explore in particular the case of an explicit symmetry breaking
induced by a constant, uniform magnetic field $B$ acting on a system of
neutrons. Here we compare calculations performed using either the
single-particle Matsubara propagator or the zero-temperature polarization
propagator, discussing under which perturbative scheme they lead to identical
results (if $B$ is not too large).
We finally address the issue of the spontaneous symmetry breaking for a system
of neutrons using the technique of the anomalous propagator: in this  framework
we recover the Stoner equation and the critical values of the interaction
corresponding to a transition to a ferromagnetic phase.
\end{abstract}


\end{frontmatter}

\section{Introduction}

In this work we first discuss the problem of the Hartree-Fock (HF) mean field
for an infinite system of fermions (nucleons) at finite temperature
$T$ in the context of Matsubara theory. In particular we study under which 
conditions and how the results obtained in the Matsubara framework in the 
$T\to~0$ limit coincide with those of the $T=0$ Feynman theory.

As is well-known, the two formalisms lead to identical results {\em in the
absence of symmetry breaking}, according to the old theorem for spin one-half
fermions of Kohn, Luttinger and Ward (KLW) \cite{Koh60,Lut60}, at least as far
as the ground state energy is concerned. Here we extend the theorem to the
single-particle propagator (and hence to any one-body observable), aiming at
transparently displaying how the cancellation of the diagrams, the mechanism at
the basis of KLW, is realized and how the exact $T=0$ limit is retrieved. 

Specifically we shall assess quantitatively the magnitude of the diagrams that,
while canceling in the $T\to 0$ limit, actually substantially contribute at 
finite $T$ (not, however, in finite systems \cite{Koh60,Lut60,Neg98}).
Concerning the size of their contribution to observables like the chemical
potential and the magnetization, it turns out to be larger in the proximity of 
the Fermi temperature $T_F$ owing to the temperature dependence of the 
self-energy. The latter, in fact, displays a marked change near $T_F$
reflecting the transition from a quantal to a classical regime. 
This transition is sensitively affected by the range of the interaction: hence
in our study we employ both a finite and a zero-range force, however of
schematic nature since we have no pretense of performing a realistic
calculation, but rather we aim to investigate the generic temperature behavior
of observables that are relevant for the finite temperature physics (such as
the chemical potential and the magnetization) in the absence or presence of a
symmetry breaking.   

We address this last issue in the second part of the paper. When a symmetry is
broken then KLW no longer holds and accordingly we explore the impact on
the HF field of such an occurrence, both when the breaking is induced
explicitly by an external field (as an example we shall consider the case of a 
constant, uniform magnetic field acting on a system of neutrons) and when it is spontaneous. 
In the first instance we show how the diagrams that would cancel each other at
small temperature in a situation of pure symmetry no longer do so: actually
their contribution grows, at $T\to 0$, keeps the value attained at $T=T_F$ if
the applied magnetic field is large enough. 

Moreover, and interestingly, in this situation the successive iterations
approach the self-consistent mean field solution smoothly or oscillating around
the HF value depending upon the \textit{sign}
of the two-body interaction among neutrons. A ferromagnetic 
force gently leads to the HF mean field, whereas an antiferromagnetic force 
entails an approach to the latter oscillating around its value, the magnitude
of the oscillations increasing with the strength of the interaction. This
behavior contrasts the one occurring in the absence of symmetry breaking where
the HF solution is always smoothly reached and, in fact, it is also rapidly
reached --- at least for a pure exchange interaction (the one we confine
ourselves to consider) ---at small and large temperature: in the first case
because of the KLW theorem, in the second because the large T domain is where
quantum physics is gone and classical physics sets in.

The unusual property, referred to above, of the finite temperature HF solutions
in presence of a magnetic field $B$ stems in part from the remarkable
occurrence that $B$, in breaking the spin rotational symmetry of the system,
induces in the interaction matrix element a direct term which otherwise would
be absent, the force we employ being of pure exchange character. This is of  
importance when one aims to recover the vanishing temperature limit of the 
observables, say the magnetization, computed in the Matsubara HF formalism in
the framework of the $T=0$ Feynman theory. This in fact turns out to be
possible, \textit{in spite of the breakdown of the KLW theorem}, using
linear response theory, whose applicability is however limited to a specific 
range of $B$, which we are able to assess \textit{quantitatively}. 

Concerning instead the case of spontaneous symmetry breaking, the
non-relativistic quantum field theory describes its occurrence at $T = 0$
through the linearization of the equations of motion, i.e. at the mean field 
level, if the interaction is strong enough \cite{Hua98}. 
This approach coincides with the many-body Hartree mean field theory at $T=0$,
which indeed exhibits a spontaneous breaking of symmetry --- again for a force
strong enough --- provided the variational search for the
determinant yielding the minimum energy is allowed to span all of the possible
symmetry configurations (we refer to this approach as the \textit{generalized
$T=0$ HF}). This is the path we follow in the last Section in the framework of
the anomalous propagator technique \cite{Mat68} at $T=0$.

Likewise, also the temperature HF theory displays a spontaneous symmetry
breaking. We prefer, however, to search for the onset of the latter in 
the framework of the $T=0$ linear response theory which, in parallel to
the temperature HF, predicts, either in ring or in the random phase
approximation (RPA), a divergent magnetic 
susceptibility of the Fermi system when the strength of the nucleon-nucleon
interaction, \textit{assumed to be ferromagnetic}, reaches a critical value
$V_{\mathrm{crit}}$, thus signaling the occurrence of a phase transition. 
For a contact interaction, when the response of the system is computed in ring
approximation, the value of $V_{\mathrm{crit}}$ thus found is identical to the
one obtained in \cite{Hua98} or in the generalized Hartree approximation,
whereas the RPA response yields a lower value (by a factor 2/3), which
coincides with the one predicted by the $T=0$ generalized HF theory. As
mentioned above, the same result is obtained using the thermal single-particle
HF propagator, computed with a contact interaction, in the $T\to0$ limit.

When the symmetry is spontaneously broken, one would like to describe the
associated Goldstone modes. 
In connection with the collective excitations of the system, at zero
temperature it turns out that, for an anti-ferromagnetic coupling among the
constituents, the Fermi system supports 
the existence of zero-sound collective modes characterized by a linear
relationship between energy and momentum. These modes correspond to the spin
waves (also referred to as magnons) found in the antiferromagnetic materials
and \textit{lower} the susceptibility. 
They represent the Goldstone modes, because our system, although without any
cristal structure, still lives in a condensed phase given by the
antiferromagnetic order of the spin constituents: we accordingly view the Fermi
gas as paramagnetic at finite $T$ and antiferromagnetic at $T=0$.

On the other hand a
ferromagnetic interaction only supports the existence of modes embedded in the
particle-hole continuum for $V<V_{\mathrm{crit}}$. In this regime such a force 
softens the response function of the system until, when 
$V=V_{\mathrm{crit}}$, the latter diverges at zero excitation energy as the
transferred momentum becomes very small.
Thus the phase transition is also displayed by the behavior of the response
function, which basically corresponds to the spin-spin correlation function. 
On the other hand, for a ferromagnetic coupling $V>V_{\mathrm{crit}}$ 
new collective modes appear, in the direction
orthogonal to the one set by the spontaneous magnetization of the system. 
These also represent Goldstone modes and will be addressed in a forthcoming
paper.

\section{Hartree-Fock theory at zero and finite temperature}

In this Section we discuss the mean-field HF theory of a system of nucleons
at finite temperature. 
For simplicity we confine ourselves to consider an infinite homogeneous system
and a schematic static exchange interaction in the non-relativistic limit. 

\begin{figure}
\begin{center}
\includegraphics[clip,height=4cm]{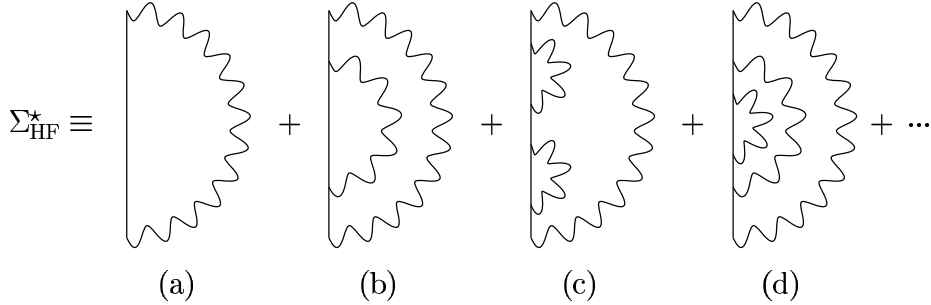}
\caption{\label{fig:HFself}The perturbative expansion of the HF
  irreducible self-energy for a pure exchange interaction.} 
\end{center}
\end{figure}
In this situation it is well-known that at zero temperature the HF mean-field
theory reduces to first order perturbation theory.
Indeed for the single-particle wave functions self-consistency is immediately
achieved in the first iteration of the HF equations owing to the translational
invariance of the system. 
In conformity, the diagrams contributing to the HF self-energy (displayed
in Fig.~\ref{fig:HFself}) vanish, except for the first one: indeed 
all of them display in the energy variable poles of order $\geq 2$ with zero
residue. 

For example, considering the diagram (b) of Fig.~\ref{fig:HFself} one has
\begin{eqnarray}
  \Sigma^{\star}_{(2)}(\vec{k}) &=& \left(\frac{i}{\hbar}\right)^{2}
    \lim_{\eta \to 0^{+}} \int \frac{{\d}\vec{k}_{1}}{(2\pi)^{3}}
    \int \frac{{\d}\vec{k}_{2}}{(2\pi)^{3}} \int_{-\infty}^{\infty}
    \frac{{\d} \omega_{1}}{2\pi} \int_{-\infty}^{\infty}
    \frac{{\d} \omega_{2}}{2\pi} e^{i\omega_{1}\eta}
    e^{i\omega_{2}\eta} \nonumber  \\
  & & \times V(\vec{k}-\vec{k}_{1}) V(\vec{k}_{1}-\vec{k}_{2}) 
    [G^{0}(\vec{k}_{1},\omega_1)]^{2} G^{0}(\vec{k}_{2},\omega_2),
\end{eqnarray}
$V$ being the interaction and $G^0$ the zero-order propagator in
a Fermi gas with Fermi momentum $k_F$. Now, in the frequency integral,
\begin{eqnarray}
  I &=& \int_{-\infty}^{\infty} \frac{{\d} \omega_{1}}{2\pi}
    e^{i\omega_{1}\eta} [G^{0}(\vec{k}_{1},\omega_1)]^{2} \nonumber \\
  &=& \int_{-\infty}^{\infty} \frac{{\d}
    \omega_{1}}{2\pi} e^{i\omega_{1}\eta}\left[
  \frac{\theta(|\vec{k}_{1}|-k_{F})}{(\omega_{1}-\omega_{\vec{k}_{1}}+i\eta)^2}   +\frac{\theta^(k_F-|\vec{k}_{1}|)}{(\omega_{1}-\omega_{\vec{k}_{1}}-i\eta)^2}
    \right] ,
\end{eqnarray}
only the double pole at 
$\omega_1=\omega_{\vec{k}_{1}}+i\eta\equiv\hbar\vec{k}_{1}^2/2m+i\eta$ 
should be considered, the contour of the integration lying in the complex upper
plane $\mathrm{Im}~\omega_{1}>0$. 
Hence one gets
\begin{equation}
  I=-\eta~ e^{i(\omega_{\vec{k}_{1}}+i\eta)\eta}\theta(k_{F}-|\vec{k}_{1}|)
  \rightarrow 0 \ \ \mathrm{for} \ \ \eta \rightarrow 0 
\end{equation} 
and likewise for all the other diagrams of Fig.~\ref{fig:HFself} of order
$\geq 2$. 

On the other hand, at finite temperature the Matsubara diagrams of
Fig.~\ref{fig:HFself} no longer vanish. In fact, taking again as an 
example the second order self-energy, one has
\begin{eqnarray}
  \Sigma^{\star}_{(2)}(\vec{k},T) &=& \left(\frac{-1}{\hbar}\right)^{2}
    \int\frac{{\d}\vec{k}_{1}}{(2\pi)^{3}} 
    \int\frac{{\d}\vec{k}_{2}}{(2\pi)^{3}}
    \frac{1}{(\beta \hbar)^{2}} \sum_{n_{1}} \sum_{n_{2}} 
    e^{i\omega_{n_{1}}\eta} e^{i\omega_{n_{2}}\eta} \nonumber \\
  && \times 
    V(\vec{k}-\vec{k}_{1}) V(\vec{k}_{1}-\vec{k}_{2}) 
    [{\mathcal{G}}^{0}(\vec{k}_{1},\omega_{n_1})]^{2} 
    {\mathcal{G}}^{0}(\vec{k}_{2},\omega_{n_2}),
\end{eqnarray}
where
\begin{equation}
  \mathcal{G}^0(\vec{k},\omega_n) = 
    \frac{1}{i\omega_n-(\epsilon_{\vec{k}}^{(0)}-\mu)/\hbar}
\end{equation}
is the thermal free propagator at finite $T=1/k_B\beta$ ($k_B$ being the
Boltzmann's constant and $\mu$ the free chemical potential). Evaluating the 
frequency sums according to the standard rule 
\begin{equation}
  \lim_{\eta \to 0^{+}} \sum_{n_{2}} e^{i\omega_{n_{2}}\eta} 
    {\mathcal{G}}^{0}(\vec{k}_{2},\omega_{n_2}) = \lim_{\eta \to 0^{+}}
    \sum_{n_{2}} \frac{e^{i\omega_{n_{2}}\eta}}{i\omega_{n_{2}}-
    (\epsilon^{(0)}_{\vec{k}_{2}}-\mu)/\hbar} =
    \frac{\beta\hbar}{e^{\beta(\epsilon^{(0)}_{\vec{k}_{2}}-\mu)}+1} 
\end{equation}
and
\begin{eqnarray}
\label{sommafrequenze}
  \lim_{\eta \to 0^{+}} \sum_{n_{1}} e^{i\omega_{n_{1}}\eta} 
    [{\mathcal{G}}^{0}(\vec{k}_{1},\omega_{n_1})]^{2} &=& \lim_{\eta \to 0^{+}}
    \sum_{n_{1}} \frac{e^{i\omega_{n_{1}}\eta}}{\left[i\omega_{n_{1}}-
    (\epsilon^{(0)}_{\vec{k}_{1}}-{\mu})/\hbar\right]^{2}} \nonumber \\
  &=&
    -\frac{(\beta\hbar)^{2} e^{\beta(\epsilon^{(0)}_{\vec{k}_{1}}-{\mu})}}
    {\left[e^{\beta(\epsilon^{(0)}_{\vec{k}_{1}}-{\mu})}+1\right]^{2}}, 
\end{eqnarray}
where $\omega_{n}=(2n+1)\pi/\beta\hbar$ and the sums run over all the positive
and negative integers, one gets 
\begin{eqnarray}
\label{secondogenerale}
  \Sigma^{\star}_{(2)}(\vec{k},T) &=& \frac{-\beta}{\hbar}
    \int \frac{{\d} \vec{k}_{1}}{(2\pi)^{3}}
    \int \frac{{\d} \vec{k}_{2}}{(2\pi)^{3}} 
    V(\vec{k}-\vec{k}_{1}) V(\vec{k}_{1}-\vec{k}_{2}) \nonumber \\
  && \times 
    \frac{e^{\beta(\epsilon^{(0)}_{\vec{k}_{1}}-{\mu})}}
    {\left[e^{\beta(\epsilon^{(0)}_{\vec{k}_{1}}-{\mu})}+1\right]^{2}} 
    \frac{1}{e^{\beta(\epsilon^{(0)}_{\vec{k}_{2}}-\mu)}+1}.
\end{eqnarray}
The self-energy (\ref{secondogenerale}) does not vanish: as a consequence, at
finite $T$, even for an infinite system, the HF equations are not trivial.

We briefly illustrate this point using, as an example, a simple
static spin-isospin dependent nucleon-nucleon central interaction of the type 
\begin{equation} 
\label{interazione}
  V = \vec{\sigma}_{1} \cdot \vec{\sigma}_{2} \vec{\tau}_{1} \cdot
    \vec{\tau}_{2} v({\vec{x}}_1-{\vec{x}}_2)
\end{equation}
which, of course, has no pretense of being realistic.

Since the system is homogeneous, the eigenfunctions are still plane waves,
whereas the HF eigenvalues read  
\begin{equation}
  \epsilon_{\vec{k}} = 
    \epsilon_{\vec{k}}^{(0)}+\hbar\Sigma^{\star}_{\mathrm{HF}}(\vec{k},T),
\end{equation}
where
\begin{equation}
  \hbar\Sigma^{\star}_{\mathrm{HF}}(\vec{k},T) = 
    - 9\int\frac{{\d}\vec{k}'}{(2\pi)^{3}} n_{\vec{k}'} v(\vec{k}-\vec{k}') 
\end{equation}
is the HF irreducible self energy for the interaction (\ref{interazione}),
\begin{equation}
  n_{\vec{k}} = \frac{1}{e^{\beta(\epsilon_{\vec{k}}-\mu)}+1}
\end{equation}
the Fermi distribution and $v(\vec{k})$ the Fourier transform of $v(\vec{x})$.

Thus the HF equations at finite temperature, unlike the case at $T=0$,
represent a non trivial self-consistency problem, because the single-particle
energies (the eigenvalues) enter not only in the equations, but in the Fermi
distribution as well. 

Accordingly, the system of equations
\begin{subequations}
\label{sistema}
\begin{eqnarray} 
    \epsilon_{\vec{k}} &=& \epsilon_{\vec{k}}^{(0)} +
    \hbar\Sigma^{\star}_{\mathrm{HF}}(\vec{k},T) 
\label{sistemaa}
    \\
    \rho &=&4\displaystyle{\int\frac{{\d}\vec{k}'}{(2\pi)^{3}} n_{\vec{k}'}},
\label{sistemab}
\end{eqnarray}
\end{subequations}
where the density $\rho$ of the system is assumed to be fixed, should be
solved. 
We have accomplished this task through numerical iterations of the above
equations (the unknowns being $\epsilon_{\vec{k}}$ and $\mu$) until reaching
self-consistency (see Subsection~\ref{subsec:HFres} for the numerical
results). 

At zero order $\epsilon_{\vec{k}}\equiv\epsilon_{\vec{k}}^{(0)}$ and one fixes 
$\mu\equiv\mu^{(0)}$ through Eq.~(\ref{sistemab}), i.~e.
\begin{equation}
\label{eq:rhomu0}
  \rho = 4 \int\frac{{\d} \vec{k}'}{(2\pi)^{3}}
    \frac{1}{e^{\beta(\epsilon_{\vec{k}'}^{(0)}-\mu^{(0)})}+1}. 
\end{equation}
At first order, from (\ref{sistemaa}) one has 
\begin{equation}
  \epsilon_{\vec{k}}^{(1)} = \epsilon_{\vec{k}}^{(0)} -
    9 \int\frac{{\d} \vec{k}'}{(2\pi)^{3}} n_{\vec{k}'}^{(0)}
    v(\vec{k}-\vec{k}') ,
\end{equation}
with $n_{\vec{k}'}^{(0)}=1/\exp[\beta(\epsilon_{\vec{k}'}^{(0)}-\mu^{(0)})+1]$;
from (\ref{sistemab}) one then has
\begin{equation}
\label{eq:rhomu1}
  \rho = 4 \int\frac{{\d} \vec{k}'}{(2\pi)^{3}} 
    \frac{1}{e^{\beta(\epsilon_{\vec{k}'}^{(1)}-\mu^{(1)})}+1} ,
\end{equation}
which fixes a new chemical potential $\mu^{(1)}$.
The latter is then put back in (\ref{sistemaa}) together with
$\epsilon_{\vec{k}}^{(1)}$ in order to generate $\epsilon_{\vec{k}}^{(2)}$ and
so on. 

Numerically the procedure is quite straightforward.
However, it is of interest to analyze its diagrammatic content,
since it helps to understand the different roles played by the various classes
of diagrams at zero and finite temperature.

\begin{figure}
\begin{center}
\includegraphics[clip,height=8cm]{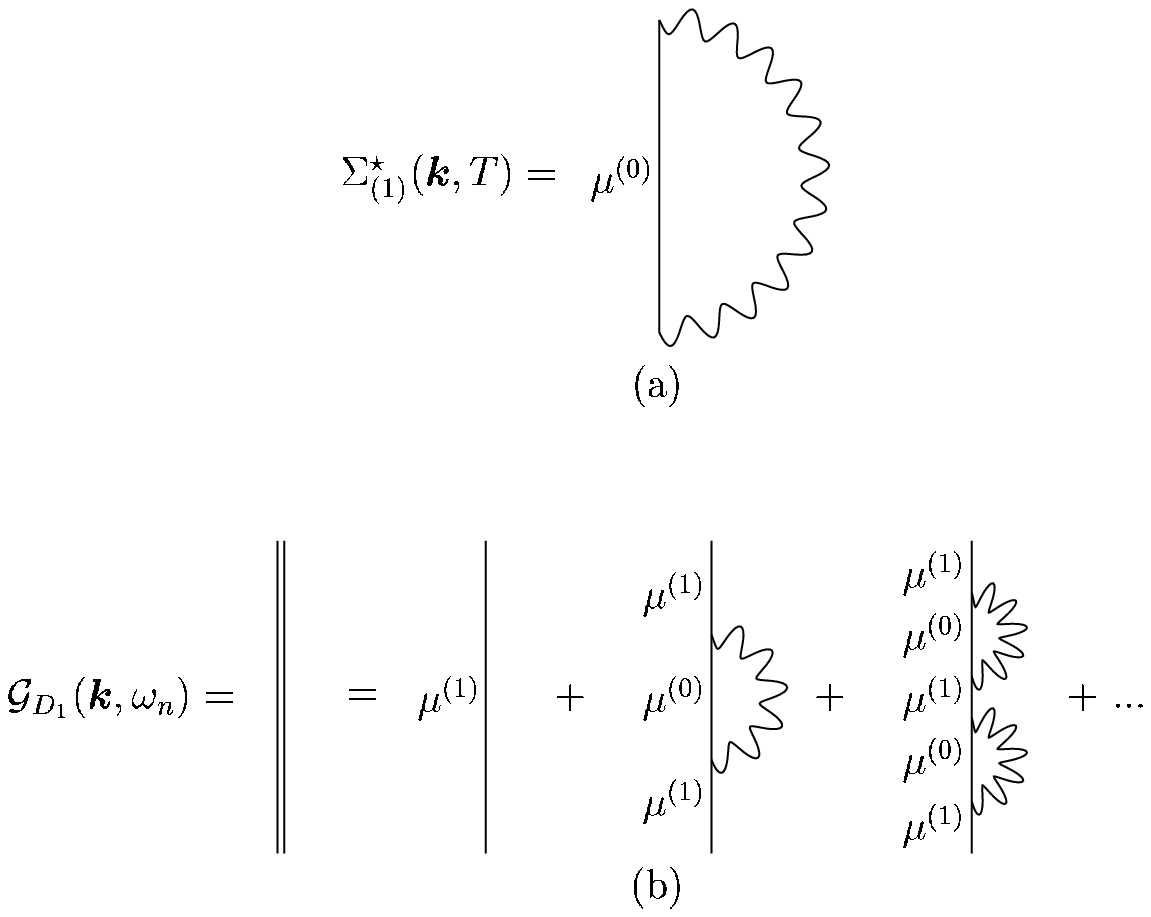}
\caption{\label{fig:GD1}The thermal first order self-energy (panel a) and the
  perturbative expansion of the thermal first-iteration HF propagator (panel
  b). } 
\end{center}
\end{figure}

Thus --- following the steps outlined above for the solution of the HF
equations --- the first order self-energy $\Sigma_{(1)}^{\star}$ is displayed
in Fig.~\ref{fig:GD1}a: it embodies $\mu^{(0)}$, the chemical potential for the
non-interacting system. 
$\Sigma_{(1)}^{\star}$, together with $\mu^{(1)}$ --- the new chemical
potential determined by the requirement of fixed density,
Eq.~(\ref{eq:rhomu1}) --- is then inserted into the Dyson equation for the
thermal propagator, namely
\begin{equation} 
\label{dyson}
  \mathcal{G}(\vec{k},\omega_n) = \mathcal{G}^{0}(\vec{k},\omega_n) +
    \mathcal{G}^{0}(\vec{k},\omega_n) \Sigma^{\star}(\vec{k},T)
    \mathcal{G}(\vec{k},\omega_n) 
\end{equation}
whose solution is (Fig.~\ref{fig:GD1}b)
\begin{equation} 
\label{paperino}
  \mathcal{G}_{D_1}(\vec{k},\omega_n) = 
    \frac{1}{i\omega_n-(\epsilon_{\vec{k}}^{(0)}+
    \hbar\Sigma^{\star}_{(1)}(\vec{k},T)-\mu^{(1)})/\hbar}.
\end{equation}

Fig.~\ref{fig:GD1}b clearly illustrates that the chemical potential
associated with the external propagation lines or with those linking different
self-energy insertions should now not be $\mu^{(0)}$, but rather $\mu^{(1)}$,
in order to keep the density of the system constant. 

\begin{figure}
\begin{center}
\includegraphics[clip,height=8cm]{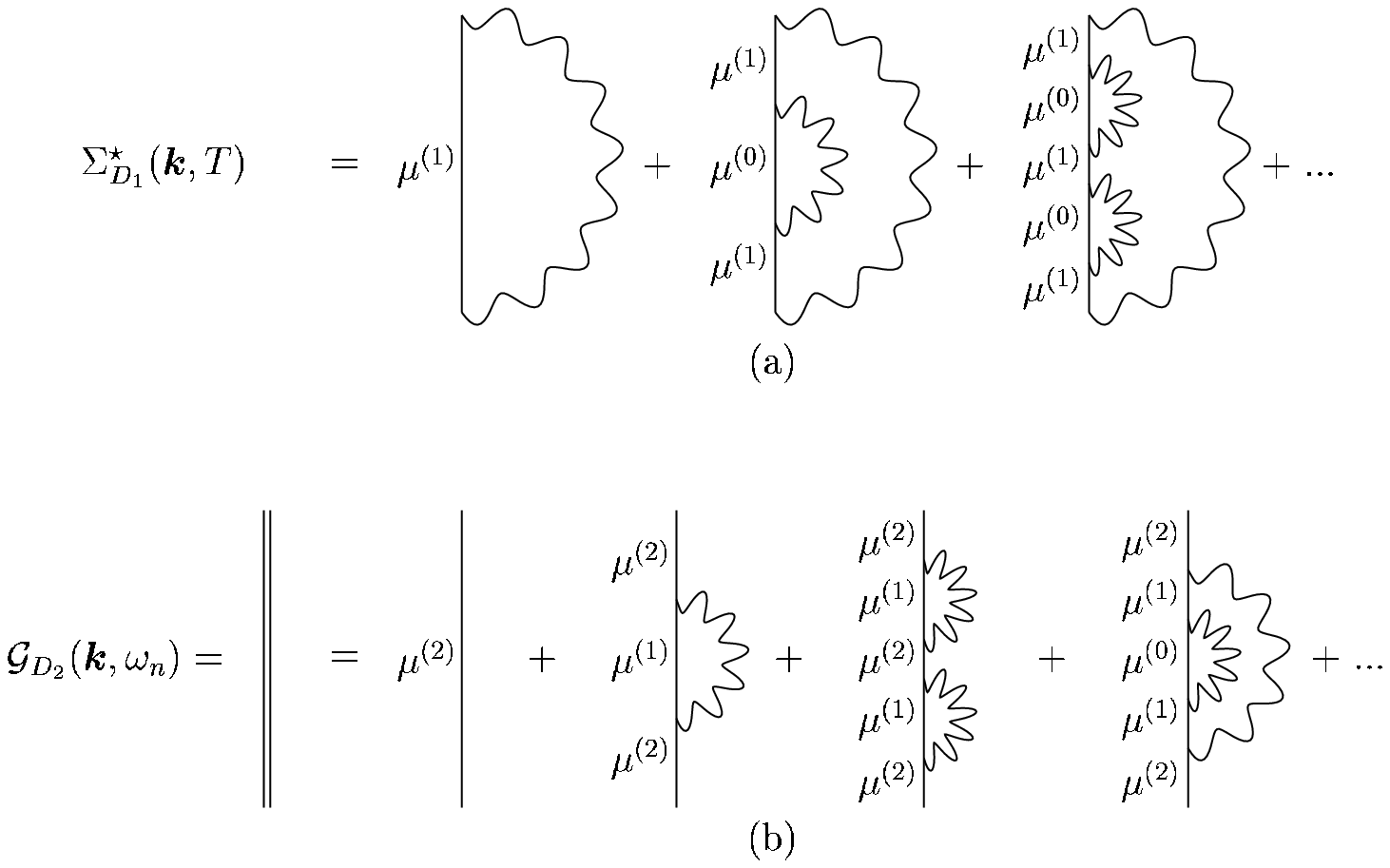}
\caption{\label{fig:GD2}Perturbative expansion of the first iteration
  irreducible HF thermal self-energy (panel a) and of the second-iteration HF
 thermal propagator (panel b).}
\end{center}
\end{figure}

One then computes with the propagator (\ref{paperino}) the new self-energy
$\Sigma^{\star}_{D_1}(\vec{k},T)$, which will embody the HF diagrams shown in
Fig.~\ref{fig:GD2}a and will be used to fix a new chemical potential
$\mu^{(2)}$. 

Then, solving the Dyson equation with $\Sigma^{\star}_{D_1}(\vec{k},T)$ as a 
kernel will obviously lead to the new propagator shown in Fig.~\ref{fig:GD2}b.
The procedure should be iterated until self-consistency is reached.
Through this iterative procedure \textit{all} the HF diagrams are generated and
naturally organized in different classes. 

\subsection{The case of a zero-range interaction}

As a preliminary we address the HF problem at finite $T$ considering the simple
case of a zero-range interaction of pure exchange nature, namely we set in
(\ref{interazione}) 
\begin{equation}
  v(\vec{x}_1 -\vec{x}_2 )= V_{1} \delta(\vec{x}_1 -\vec{x}_2 ),
\end{equation}
whose Fourier transform is of course just $V_{1}$.

The HF equations (\ref{sistema}) in this case reduce to:
\begin{equation}
\label{pippo}
  \left\{\begin{array}{rl}
    \epsilon_{\vec{k}} & = \epsilon_{\vec{k}}^{(0)} 
    -9 V_{1} \displaystyle{\int\frac{{\d}
    \vec{k}'}{(2\pi)^3}}\frac{1}{e^{\beta(\epsilon_{\vec{k}'}-\mu)}+1} \\ 
    \rho & = 4 \displaystyle{\int\frac{{\d} \vec{k}'}{(2\pi)^{3}}}
    \frac{1}{e^{\beta(\epsilon_{\vec{k}'}-\mu)}+1}. 
  \end{array}\right.
\end{equation}
At zero-order $\epsilon_{\vec{k}} = \epsilon_{\vec{k}}^{(0)}$
and the chemical potential $\mu^{(0)}$ (a few terms of its expansion as a 
function of $k_F$ and $T$ are quoted in Appendix A) is fixed by the requirement
of yielding  the correct density (see Eq.~(\ref{eq:rhomu0})).

By inserting these results into Eq.~(\ref{pippo}) we get
\begin{eqnarray}
  \epsilon_{\vec{k}}^{(1)} &=&
    \epsilon_{\vec{k}}^{(0)}+\hbar\Sigma^{\star}_{(1)} 
    \nonumber \\
  &=& \epsilon_{\vec{k}}^{(0)} - 9 V_{1} \int\frac{{\d} \vec{k}'}{(2\pi)^3}
    \frac{1}{e^{\beta(\epsilon_{\vec{k}'}^{(0)}-\mu^{(0)})}+1} \nonumber \\ 
  &=& \epsilon_{\vec{k}}^{(0)} -\frac{9}{4} V_{1} \rho.
\end{eqnarray}
Again, the chemical potential has now to be redefined in order
to keep the density constant (see Eq.~(\ref{eq:rhomu1})).
It is immediately seen that in the present case of a zero-range interaction
the change of $\mu$ amounts to a mere constant shift, i.e. 
\begin{equation}
\label{pluto}
  \mu^{(1)} = \mu^{(0)}-\frac{9}{4} V_{1} \rho .
\end{equation}
Performing then a second iteration we obtain
\begin{eqnarray}
\label{pinco}
  \epsilon_{\vec{k}}^{(2)} &=& \epsilon_{\vec{k}}^{(0)} +
    \hbar\Sigma^{\star}_{D_1}(\vec{k}) \nonumber \\
  &=&\epsilon_{\vec{k}}^{(0)} -9 V_{1}
    \int\frac{{\d} \vec{k}'}{(2\pi)^3}
    \frac{1}{e^{\beta(\epsilon_{\vec{k}'}^{(1)}-\mu^{(1)})}+1} \nonumber \\
  &=&\epsilon_{\vec{k}}^{(0)}-\frac{9}{4} V_{1}\rho = \epsilon_{\vec{k}}^{(1)},
\end{eqnarray}
where $\hbar\Sigma^{\star}_{D_1}$ is here computed using the
propagator (\ref{paperino}), which, for a zero-range interaction, reads 
\begin{equation}
  \mathcal{G}_{D_1}(\vec{k},\omega_n) =
    \frac{1}{i\omega_n-(\epsilon_{\vec{k}}^{(0)} -
    9V_{1}\rho/4-\mu^{(1)})/\hbar}. 
\end{equation}
Since Eq.~(\ref{pinco}) clearly entails the identity
\begin{equation}
\label{lorena}
  \hbar\Sigma^{\star}_{(1)} = \hbar\Sigma^{\star}_{D_1},
\end{equation}
without any depence on $T$, we conclude that self-consistency is immediately
achieved at any temperature for a zero-range force. In other words, for such an
interaction the HF problem is trivial both at $T = 0$ and at finite $T$.
  
\begin{figure}
\begin{center}
\includegraphics[clip,height=4cm]{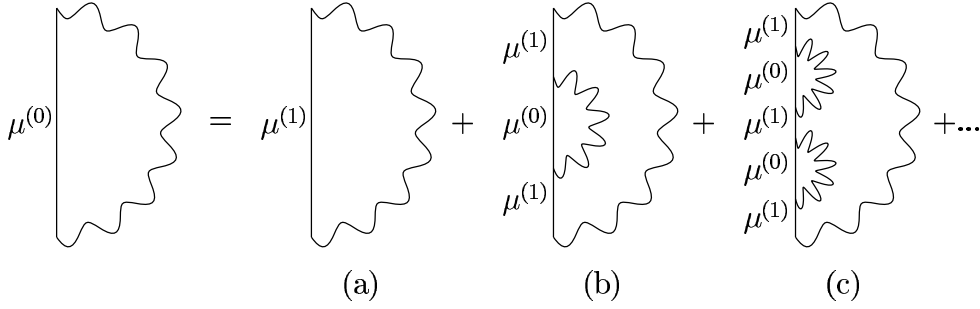}
\caption{\label{fig:Sigma1SigmaD1}Diagrammatic representation of the equation 
  $\Sigma^{\star}_{(1)}=\Sigma^{\star}_{D_1}$}.
\end{center}
\end{figure}

The diagrammatic content of Eq.~(\ref{lorena}), displayed in
Fig.~\ref{fig:Sigma1SigmaD1}, helps in understanding this conclusion:
in fact, Eq.~(\ref{lorena}) holds valid because the diagrams on the right hand
side of Fig.~\ref{fig:Sigma1SigmaD1} \textit{cancel at any temperature order by
order in the coupling constant $V_1$}. Note that $V_1$ appears both in the
interaction lines and in the chemical potential $\mu^{(1)}$: hence even the
diagram (a) in Fig.~\ref{fig:Sigma1SigmaD1} contributes to all orders in $V_1$. 

To illustrate analytically these cancellations we evaluate diagram (a) to the 
order $O(V_1^3)$ using Eq.~(\ref{pluto}). We get
\begin{eqnarray}
\label{topolino}
  \Sigma^{\star(a)}_{D_1} &=& - 9\frac{V_{1}}{\hbar}
    \int\frac{{\d} \vec{k}}{(2\pi)^3}
    \frac{1}{e^{\beta(\epsilon_{\vec{k}}^{(0)}-\mu^{(1)})}+1} \nonumber \\ 
  &=& - 9\frac{V_{1}}{\hbar} \int\frac{{\d} \vec{k}}{(2\pi)^3}
    \frac{1}{e^{\beta(\epsilon_{\vec{k}}^{(0)}-\mu^{(0)})}+1}
    e^{-9\beta V_1\rho/4} \nonumber \\
  && \qquad \times \left[ 1+ \frac{\sum_{n=1}^{\infty}
    (-9\beta V_1\rho/4)^n/n!}
    {e^{\beta(\epsilon_{\vec{k}}^{(0)}-\mu^{(0)})}+1}\right]^{-1} \nonumber \\ 
  &=& - \frac{9}{4}\frac{V_{1}}{\hbar}\rho +
    (9V_1)^2 \frac{\beta\rho}{4\hbar}
    \int\frac{{\d} \vec{k}}{(2\pi)^3}
    \frac{e^{\beta(\epsilon_{\vec{k}}^{(0)}-\mu^{(0)})}}
    {\left[e^{\beta(\epsilon_{\vec{k}}^{(0)}-\mu^{(0)})}+1\right]^2} +
    O\left(V_1^3\right). 
\end{eqnarray}
From the above one sees that on the right hand side of Eq.~(\ref{lorena}) the
term linear in 
$V_1$ cancels with the left hand side, whereas the quadratic term cancels with
the leading order term in $V_1$ associated with the diagram (b). Indeed,
computing the latter at the lowest order, that is with $\mu^{(0)}$ in all the 
propagation lines, one obtains \begin{eqnarray}
  \Sigma^{\star(b)}_{D_1} &=&
    (9V_1)^2\left(-\frac{1}{\beta\hbar^2}\right)^2\int\frac{{\d} 
    \vec{k}}{(2\pi)^3}\int\frac{{\d} \vec{k}'}{(2\pi)^3}
    \sum_{n'=-\infty}^{\infty}\frac{e^{i\omega_{n'}\eta'}}
    {i\omega_{n'}-(\epsilon_{\vec{k}'}^{(0)}-\mu^{(0)})/\hbar} \nonumber \\
  && \quad \times \sum_{n=-\infty}^{\infty}\frac{e^{i\omega_{n}\eta}}
    {\left(i\omega_{n}-(\epsilon_{\vec{k}}^{(0)}-\mu^{(0)})/\hbar\right)^2} +
    O\left(V_1^3\right).
\end{eqnarray}
Performing then the sums over the Matsubara frequencies with the usual contour
integration techniques as in Eq.~(\ref{sommafrequenze}), one gets a
contribution 
precisely canceling the term quadratic in $V_1$ in Eq.~(\ref{topolino}). 
By the same arguments the cancellation is proved to hold at any order in
$V_1$. In the following we shall refer to this occurrence equivalently as the
\textit{cancellation theorem} or the KLW theorem.

\subsection{The case of a finite-range interaction} 
\label{subsec:rangefinito}

We now show that the cancellation mechanism previously illustrated occurs
(under certain conditions of symmetry) also for a finite range potential, but,
strictly speaking, in the limit $T\to 0$ only. 

For this purpose let 
\begin{equation}
\label{finito}
  v({\vec{x}}_1-{\vec{x}}_2) = V_{1}\frac{\lambda^{2}}{4\pi}
    \frac{e^{-\lambda |{\vec{x}}_1-{\vec{x}}_2|}}{|{\vec{x}}_1-{\vec{x}}_2|},
\end{equation}
which yields back the zero-range force in the $\lambda\to\infty$ limit and
whose Fourier transform is
\begin{equation}
  v(k) = V_1\frac{\lambda^2}{\lambda^2+k^2}
\end{equation}
(note that in the figures we shall express the range parameter $\lambda$ in
MeV). The first order self-energy is then easily obtained and reads
\begin{equation}
\label{unofinito}
  \Sigma^{\star}_{(1)}(\vec{k},T) = 
    -9\frac{V_{1}}{\hbar}\frac{\lambda^2}{(2\pi)^2}
    \frac{1}{2k} \int_{0}^{\infty}{\d}k_{1} k_{1}\ln
    \frac{\lambda^2+(k+k_1)^2}{\lambda^2+(k-k_1)^2}
    \frac{1}{e^{\beta(\epsilon_{\vec{k}_1}^{(0)}-\mu^{(0)})}+1}. 
\end{equation}
In the $T\to 0$ limit, where the Fermi distribution reduces to a $\theta$
function, the integration in Eq.~(\ref{unofinito}) can be done analytically, 
yielding 
\begin{eqnarray}
\label{unofinitozero}
  \Sigma^{\star}_{(1)}(\vec{k},T=0) &=&
    -9\frac{V_{1}}{\hbar}\frac{\lambda^3}{(2\pi)^2}\left[
    \frac{k_F}{\lambda}-\left(\arctan\frac{k_F-k}{\lambda} +
    \arctan\frac{k_F+k}{\lambda}\right)+\right. \nonumber\\  
  && \qquad \left.+\frac{1}{4}\left(\frac{k_F^2}{\lambda k} -
    \frac{k}{\lambda}+\frac{\lambda}{k}\right)\ln
    \frac{1+(k_F+k)^2/\lambda^{2}}{1+(k_F-k)^2/\lambda^{2}}\right]. 
\end{eqnarray}
At second order one gets
\begin{eqnarray}
\label{eq:Sigmas2T}
  \Sigma^{\star}_{(2)}(\vec{k},T) &=& \frac{-\beta}{\hbar}
    \left(9V_{1}\frac{\lambda^2}{(2\pi)^2}\right)^2 \frac{1}{4k}
    \int_{0}^{\infty} {\d} k_{1} \int_{0}^{\infty} {\d} k_{2} 
    k_2 \ln\frac{\lambda^2+(k+k_1)^2}{\lambda^2+(k-k_1)^2} \nonumber\\
  && \times\ln\frac{\lambda^2+(k_1+k_2)^2}{\lambda^2+(k_1-k_2)^2}
    \frac{e^{\beta(\epsilon^{(0)}_{k_{1}}-\mu^{(1)})}}{\left(
     e^{\beta(\epsilon^{(0)}_{k_{1}}-\mu^{(1)})}+1\right)^{2}} 
    \frac{1}{e^{\beta(\epsilon^{(0)}_{k_{2}}-\mu^{(0)})}+1}.
\end{eqnarray}
The above can be obtained from Eq.~(\ref{secondogenerale}), provided one
inserts $\mu=\mu^{(0)}$ in the innermost propagator and $\mu=\mu^{(1)}$
elsewhere, according to the previous discussion (see Fig.~\ref{fig:GD2}a).

Eq.~(\ref{eq:Sigmas2T}) contains a Fermi distribution and a factor which is
proportional to the derivative of a Fermi distribution with respect to the
energy $\epsilon^{(0)}_{\vec{k}_{1}}$.
These, in the limit of $T\to 0$, yield a theta function,
$\theta(\mu^{(0)}-\epsilon^{(0)}_{\vec{k}_{2}})$, and a $\delta$ distribution,
$\delta(\mu^{(1)}-\epsilon^{(0)}_{\vec{k}_{1}})$, respectively. 
Thus also $\Sigma^{\star}_{(2)}$, in this limit, can be analytically expressed
as
\begin{eqnarray}
\label{duefinitozero}
  &&\Sigma^{\star}_{(2)}(\vec{k},T=0) = -\frac{1}{\hbar^2}
    \left(9V_{1}\frac{\lambda^2}{(2\pi)^2}\right)^2 \frac{1}{8k}
    \sqrt{\frac{m}{2\mu^{(1)}}} \nonumber \\
  && \qquad \times \ln\left[
    \frac{\lambda^2+\left(k+\sqrt{2m\mu^{(1)}}/\hbar\right)^2}
    {\lambda^2+\left(k-\sqrt{2m\mu^{(1)}}/\hbar\right)^2}\right]
    \Biggl[4\frac{\sqrt{2m\mu^{(1)}}}{\hbar}k_F 
    + 4\frac{\sqrt{2m\mu^{(1)}}}{\hbar} \nonumber \\
  && \qquad \times \lambda\left(
    \arctan\frac{\sqrt{2m\mu^{(1)}}/\hbar-k_F}{\lambda} - 
    \arctan\frac{\sqrt{2m\mu^{(1)}}/\hbar+k_F}{\lambda}\right) \nonumber\\ 
  && \qquad + \left(k_F^2-\frac{2m\mu^{(1)}}{\hbar^2}+\lambda^2\right)\ln
    \frac{\lambda^2+\left(\sqrt{2m\mu^{(1)}}/\hbar+k_F\right)^2}{\lambda^2+
    \left(\sqrt{2m\mu^{(1)}}/\hbar-k_F\right)^2}\Biggr] 
\end{eqnarray}
and is {\em not vanishing}.
Nevertheless, also for finite range forces self-consistency is immediately 
achieved for $T\to0$.

The proof proceeds along the lines followed in the case of a zero-range
interaction. Indeed, since the chemical potential at first order is given by
\begin{eqnarray}
  \mu^{(1)} &=& \mu^{(0)}+\hbar\Sigma^{\star}_{(1)}(k_F,T=0) \nonumber \\
  &=& \mu^{(0)}-9V_{1}\frac{\lambda^2k_F}{(2\pi)^2}\left(1-\frac{\lambda}{k_F}
    \arctan\frac{2k_F}{\lambda} + \frac{\ln\left(1+4k_F^2/\lambda^2\right)}
    {4k_F^2/\lambda^2}\right),
\end{eqnarray}
we can compute, through the same power expansion illustrated in 
Eq.~(\ref{topolino}), the diagram (a) of Fig.~\ref{fig:Sigma1SigmaD1} to the 
order $O\left(V_1^3\right)$, getting
\begin{eqnarray}
  &&\Sigma^{\star}_{D_1}(\vec{k},T=0) = \Sigma^{\star}_{(1)}(\vec{k},T=0) + 
    \frac{1}{\hbar^3}
    \left[9V_{1}\frac{\lambda^2}{(2\pi)^2}\right]^2 \frac{mk_F}{2k}
    \nonumber \\
  &&\quad\times \left[1-\frac{\lambda}{k_F}\arctan\frac{2k_F}{\lambda} +
    \frac{\ln\left(1+4k_F^2/\lambda^2\right)}{4k_F^2/\lambda^2}\right]
    \ln\frac{\lambda^2+(k+k_F)^2}{\lambda^2+(k-k_F)^2}+O(V_{1}^{3}). 
    \nonumber \\ 
\end{eqnarray}
We thus see that the second term on the right hand side of the above equation
exactly cancels the contribution of order $V_1^2$ arising from the diagram (b)
of Fig.~\ref{fig:Sigma1SigmaD1}, whose analytic expression is given by
Eq.~(\ref{duefinitozero}) computed with $\mu^{(0)}=\hbar^2k_F^2/2m$ and
replacing $\mu^{(1)}$ with $\mu^{(0)}$ in all places, i.e. in all the
propagation lines. This is the essence of the cancellation theorem. 

Somewhat surprisingly the validity of the latter actually extends over a
remarkably large range of temperatures, as it appears from the numerical
results reported in Subsection~\ref{subsec:HFres}.

\subsection{Numerical results}
\label{subsec:HFres}

We now present a few numerical results to assess quantitatively the role of the
higher order perturbative HF diagrams at finite $T$ and the behavior of the
chemical potential as a function of temperature and density. We also explore
the sensitivity of our outcomes to the values of the parameters characterizing
our schematic interaction.

\begin{figure}
\begin{center}
\includegraphics[clip,height=8.5cm]{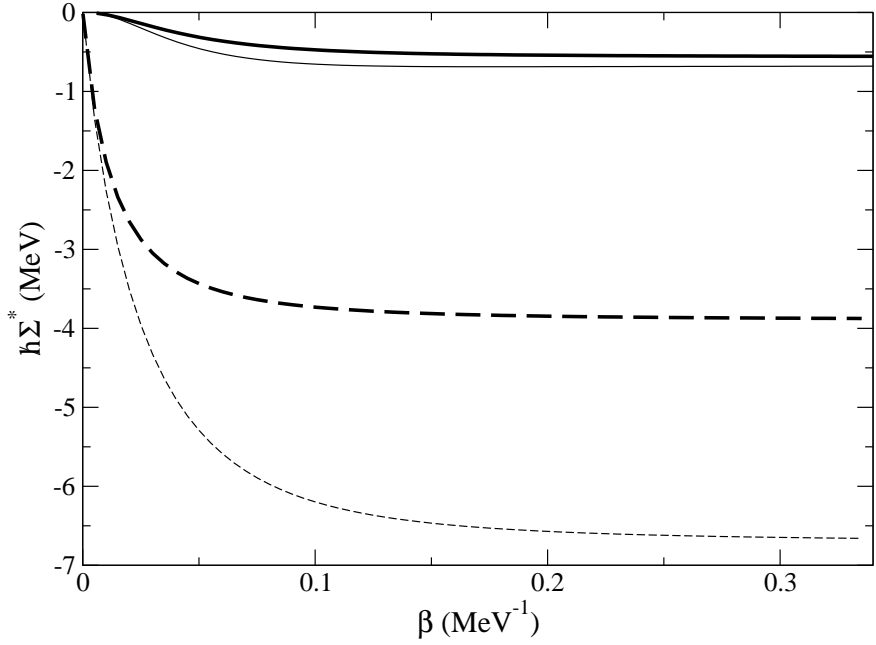}
\caption{$\hbar\Sigma^{\star}_{(1)}$ (dashed) and
  $\hbar\Sigma^{\star}_{(2)}$ (solid) as a function of $\beta$ with
  $\lambda=140$~MeV and $V_1=50$~MeV~fm$^3$ for two different
  values of $k$: $k=0$ (light lines) and $k=k_F=1.36$~fm$^-1$ (heavy lines).   
  \label{fig:selfenergybeta}} 
\end{center}
\end{figure}

\begin{figure}
\begin{center}
\includegraphics[clip,height=9cm]{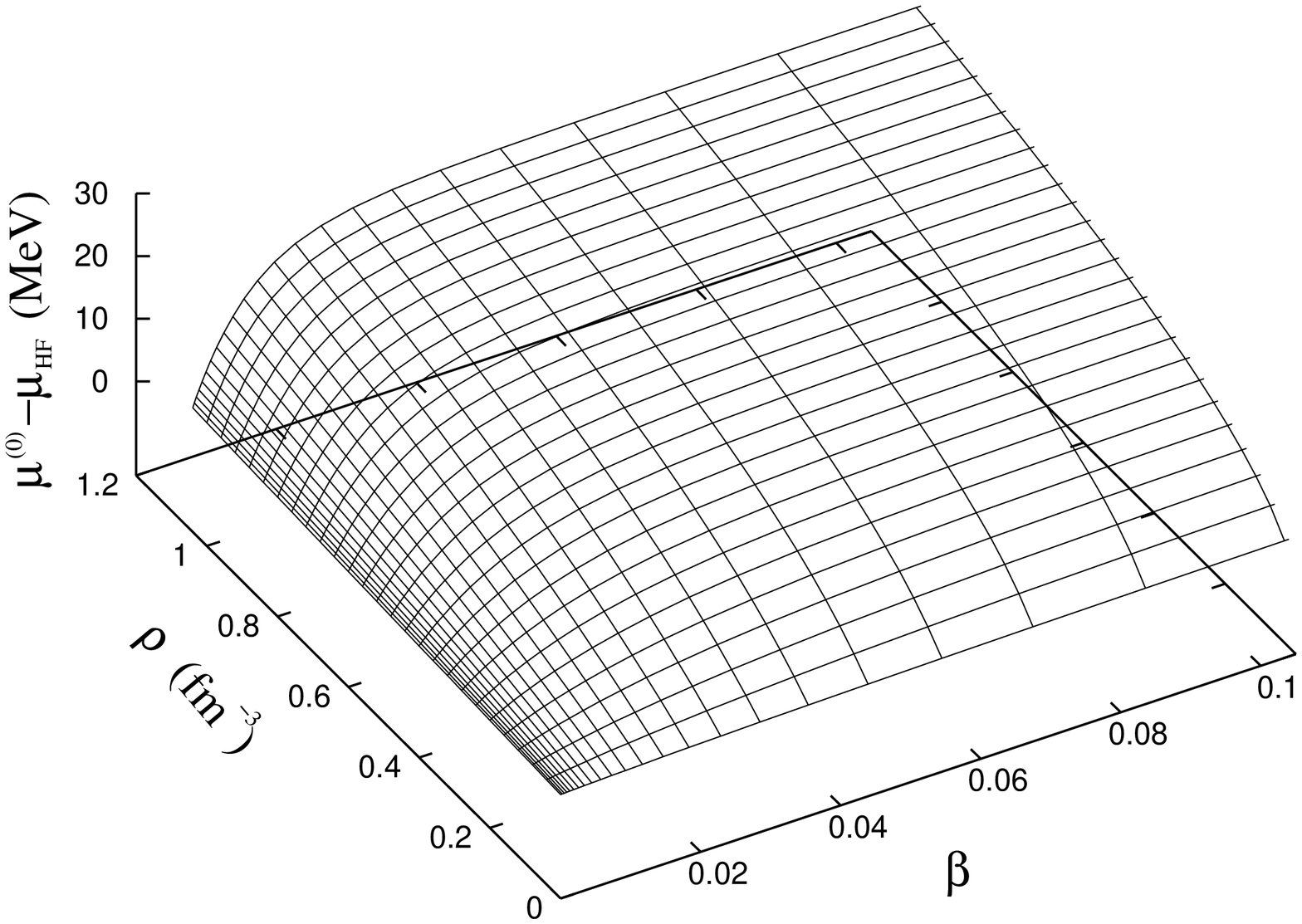}
\caption{\label{fig:muselfbetarho}$\mu^{(0)} -\mu_{\mathrm{HF}}$ as a function
  of $\beta$ and $\rho$ for a finite range force with $\lambda=140$~MeV
  and $V_1=150$~MeV~fm$^3$. }
\end{center}
\end{figure}

Specifically, we compute:
\begin{itemize}
\item[1)] the temperature behavior of the self-consistent HF chemical potential
  $\mu_{\mathrm{HF}}$ at a fixed density for various strengths and ranges of
  the force (Fig.~\ref{fig:muselfbeta}); 
\item[2)] the size of the difference $\mu^{(i)}-\mu^{(i+1)}$ between the 
  successive iterations leading to the HF chemical potential, as a function of 
  $T$ to asses the impact of the higher order perturbative terms on
  $\mu_{\mathrm{HF}}$ (Fig.~\ref{fig:pione150differ}); 
\item[3)] the temperature dependence of the first and second order
  contributions to the HF self-energy (Fig.~\ref{fig:selfenergybeta}). 
\item[4)] the temperature and density dependence of
  $\mu^{(0)}-\mu_{\mathrm{HF}}$ 
  for a given interaction (Fig.~\ref{fig:muselfbetarho}); 
\end{itemize}
\begin{figure}
\begin{center}
\includegraphics[clip,height=8.5cm]{fig_muselfbeta.eps}
\caption{\label{fig:muselfbeta} The difference $\mu^{(0)} -\mu_{\mathrm{HF}}$
  as a function of  
  $\beta$ for the cases of a force of zero-range (dotted) and finite range:
  $\lambda=3000$~MeV (dashed), $\lambda=770$~MeV (dot-dashed), 
  $\lambda=140$~MeV (solid). The light lines refer to a strength 
  $V_1=150$~MeV~fm$^3$, the heavy ones to $V_1=50$~MeV~fm$^3$. 
  The density is $\rho=0.17$ nucleons/fm$^3$.}
\vskip 1cm
\includegraphics[clip,height=8.5cm]{fig_pione150differ.eps}
\caption{\label{fig:pione150differ} The differences $\mu^{(i)}-\mu^{(i+1)}$,
  computed in HF at finite $T$, 
  as function of $\beta$ in the case of a finite range interaction with
  $\lambda=140$~MeV and $V_1=150$~MeV~fm$^3$.
  Note that the peak of the curves essentially corresponds to the Fermi
  temperature.} 
\end{center}
\end{figure}
In Fig.~\ref{fig:muselfbeta} the HF chemical potential is seen to be simply
shifted from the non-interacting value $\mu^{(0)}$ by a constant amount over a
wide range of temperature, as it happens for a zero-range interaction.

Notably this shift turns out to be essentially given by the first order proper
self-energy computed, for $T=0$, at the Fermi surface. Since
$\hbar\Sigma^{\star}_{(1)}$ turns out to be remarkably stable with the
temperature (see Fig.~\ref{fig:selfenergybeta}), we infer that the cancellation
theorem, strictly valid only in the $T\to0$ limit, actually remains operative
over a quite large range of temperatures, almost until the Fermi temperature.
This is defined as 
\begin{equation}
\label{tempfermi}
  k_BT_F=\epsilon_F=\frac{1}{\beta_F}
\end{equation}
and one has $k_BT_F=34.48$~MeV ($\beta_F=0.029$~MeV$^{-1}$) for the interaction
in Eq.~(\ref{finito}) with $\lambda=140$~MeV and $V_1=50$~MeV~fm$^3$ and
$k_BT_F=26.7$~MeV ($\beta_F=0.037$~MeV$^{-1}$) for the same interaction, but
with $V_1=150$~MeV~fm$^3$. 

We then conclude that for temperatures up to the proximity of $T_F$ the
chemical potential is
\begin{eqnarray}
\label{circa}
  \mu_{\mathrm{HF}}= \epsilon_{F}^{(0)} + \hbar\Sigma^{\star}_{(1)}(k=k_F,T=0),
\end{eqnarray}
as follows by comparing the $T\to0$ limit of the self-energy shown in 
Fig.~\ref{fig:selfenergybeta} (heavy dashed line) with the absolute value 
for large $\beta$ (3.89 MeV) of the heavy solid line 
in Fig.~\ref{fig:muselfbeta}, which expresses $\mu^{(0)}-\mu_{\mathrm{HF}}$. 

It is also evident in Fig.~\ref{fig:muselfbeta} that $\mu_{\mathrm{HF}}$ goes
to $\mu^{(0)}$ for small values of $\beta$ (classical limit), as it should,
since here all the Feynman diagrams vanish.
In fact, our exchange interaction is of purely quantum nature and, as such, is
bound to vanish at large temperature. This behavior is also transparent in
Fig.~\ref{fig:selfenergybeta}. 

The findings of Fig.~\ref{fig:muselfbeta} are complemented by those of
Fig.~\ref{fig:pione150differ}, which conveys the information on the number of
iterations required to achieve self-consistency. 
The figure shows that higher order iterations (i.e. diagrams) become
significant for $T$ approaching $T_F$, the temperature separating the classical
from the quantum regime. 
This outcome illustrates the significance of $T_F$: in fact, a
degenerate normal Fermi system, namely with a well-defined Fermi surface, lives
in the temperature range $T<T_F$. Accordingly, the impact of the Fock's
diagrams of order larger than 
one --- that tend to disrupt the Fermi surface --- become appreciable
precisely at the Fermi temperature. Away from $T_F$ their contribution is
gradually disappearing both for $T\to0$ and $T\to \infty$ but, as previously
discussed, for two radically different reasons, i.~e. the cancellation theorem
and the classical behavior (all the diagrams going to zero), respectively. 
Hence, for small $T$ the first iteration provides an almost exact estimate of
the self-energy, in spite of the fact that, as we can clearly see in 
Fig.\ref{fig:selfenergybeta}, $\hbar\Sigma^{\star}_{(2)}$, far from vanishing 
at small values of $T$, approaches the asymptotic limit given by 
Eq.~(\ref{duefinitozero}).

In concluding this Section we offer a global view of the behavior of
$\mu^{(0)} - \mu_{\mathrm{HF}}$ as a function of both $\beta$ and $\rho$ in
Fig.~\ref{fig:muselfbetarho}, where the features of the chemical potential
above discussed clearly appear.

\section{Propagation in presence of an induced symmetry breaking: the case of
  the external magnetic field} 

In this Section we address the problem of the relationship between the
Matsubara and Feynman theories at $T=0$ in the presence of a dynamically
induced symmetry breaking. In this situation the two theories
provide different results \cite{Neg98}, the correct ones being given by
Matsubara. We revisit this issue first for 
the well-known example of a non-interacting system of neutrons placed in an
external constant and uniform magnetic field pointing into the $z$-direction,
whose second-quantized hamiltonian reads 
\begin{equation}
  \hat{H}=\int{\d}\vec{x}\hat{\Psi}^{\dagger}_{\alpha}(\vec{x}) 
    \left[\delta_{\alpha\beta} \left(-\frac{\hbar^2\nabla^2}{2m}\right) + 
    V_{\alpha\beta}(\vec{x})\right]\hat{\Psi}_{\beta}(\vec{x}) , 
\end{equation}
where
\begin{equation}\label{chieri}
  V_{\alpha\beta} = -\mu_{0}~\vec{B}\cdot\vec{\sigma}_{\alpha\beta}=
    -\mu_{0}~B~\sigma^{z}_{\alpha\beta} ,
\end{equation}
$\alpha$ and $\beta$ being spin indices and $\mu_0$ the magnetic moment.

Next we switch on the interaction among the neutrons: in this case, as we 
shall see, interesting and, to our knowledge, new aspects of the temperature 
HF theory emerge.

In the interacting case the KLW theorem no longer holds  for the static 
susceptibility (namely the one obtained with the single-particle propagator). 
However the zero temperature polarization propagator leads to a
\textit{dynamical} susceptibility identical to the static one, obtained in the
$T\to0$ Matsubara framework, in the limit of vanishing momenta. 

\subsection{Static magnetization: the propagator for the non-interacting Fermi
  system} 

We first briefly consider the calculation of the free Fermi gas static
magnetization $\hat{M}$ through the standard formulas 
\begin{equation}\label{torino}
  \langle\hat{M}\rangle = -\frac{i}{\Omega}~\mu_{0} \int{\d}\vec{x}
    \sum_{\beta\alpha}\sigma^{z}_{\beta\alpha} 
    G_{\alpha\beta}(\vec{x}t,\vec{x}t^{+})
\end{equation}
at $T=0$, and
\begin{equation}
  \langle\hat{M}\rangle = \frac{1}{\Omega} \mu_{0}\int{\d}\vec{x}
    \sum_{\beta\alpha}\sigma^{z}_{\beta\alpha}
    \mathcal{G}_{\alpha\beta}(\vec{x}\tau,\vec{x}\tau^{+})
\end{equation}
at finite $T$, $\Omega$ being the large volume enclosing the Fermi gas.

\begin{figure}
\begin{center}
\includegraphics[clip,height=4cm]{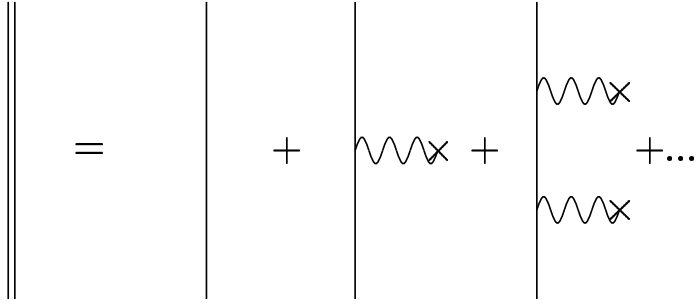}
\caption{\label{fig:dysonmagnetico}The Dyson's propagator for a non-interacting
  Fermi system in a uniform, constant magnetic field.} 
\end{center}
\end{figure}

One could think to obtain the zero-temperature propagator in Eq.~(\ref{torino})
by solving the Dyson equation with the constant proper self-energy 
\begin{equation}
  \Sigma_{\alpha\beta}^{\star} = (-1)^{\alpha+\frac{1}{2}} 
    \frac{\mu_{0}B}{\hbar},
\end{equation}
i.e. viewing the external field as a perturbation (see
Fig.~\ref{fig:dysonmagnetico}).  

One would end up with the expression
\begin{equation}\label{pino}
  G_{\alpha\beta}(\vec{k},\omega)=\delta_{\alpha\beta}
    \left[\frac{\theta(k-k_F)}{\omega-\omega_{k\alpha}+i\eta}+
    \frac{\theta(k_F-k)}{\omega-\omega_{k\alpha}-i\eta}\right] ,
\end{equation}
which is diagonal in spin space, but not proportional to the unit matrix
since 
\begin{equation}
  \omega_{k\alpha}=\frac{\hbar
  k^2}{2m}+(-1)^{\alpha+\frac{1}{2}}\frac{\mu_{0}B}{\hbar}. 
\end{equation}
In Eq.~(\ref{pino}) $k_F$ coincides with the Fermi momentum of the
non-interacting system \textit{before} switching on the magnetic field.

However, by inserting Eq.~(\ref{pino}) into Eq.~(\ref{torino}), it is
immediately found that 
\begin{eqnarray}
\label{lecce}
  <\hat{M}> & = & - \frac{i}{\Omega}\mu_0 \int{\d}\vec{x}
    \int\frac{{\d}\vec{k}}{(2\pi)^3} \int\frac{{\d}\omega}{2\pi}
    e^{i\omega\eta}\sum_{\beta\alpha}\sigma^{z}_{\beta\alpha}
    G_{\alpha\beta}(\vec{k},\omega) \nonumber \\
  &=& -i \mu_0\int\frac{{\d}\vec{k}}{(2\pi)^3}
    \int\frac{{\d}\omega}{2\pi} 
    e^{i\omega\eta}\left[\frac{\theta(k-k_F)}{\omega-\omega_{k+}+i\eta}+
    \frac{\theta(k_F-k)}{\omega-\omega_{k+}-i\eta} \right. \nonumber \\
  && \quad \left. - \frac{\theta(k-k_F)}{\omega-\omega_{k-}+i\eta} -
    \frac{\theta(k_F-k)}{\omega-\omega_{k-}-i\eta}\right]=0 ,
\end{eqnarray}
since the contributions of the two poles lying in the upper (Im$\omega >0$)
complex energy plane cancel out \cite{Neg98}.

\begin{figure}
\begin{center}
\includegraphics[clip,height=4cm]{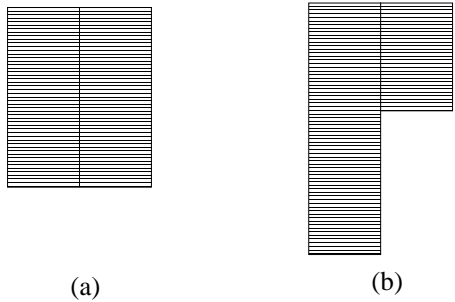}
\caption{\label{fig:energiafermi}The Fermi energy for a non-interacting Fermi
  system: in its ground state (a) and in the ground state in an external
 magnetic field (b). 
 The two boxes represent particles with spin up and down, respectively.} 
\end{center}
\end{figure}

This result is obviously wrong, since the propagator in Eq.~(\ref{pino})
(diagrammatically displayed in Fig.~(\ref{fig:dysonmagnetico})) does not 
correspond to an equilibrium state of the system (i.e. it does not 
correspond to a minimum of the energy): the true ground state of the system
(see Fig.~\ref{fig:energiafermi}) is obviously unreachable perturbatively,
since the vertex in Eq.~(\ref{chieri}) cannot flip the spin of the
constituents. 

The situation is different at finite $T$. Here the \textit{exact} Matsubara
propagator reads
\begin{equation}
  \mathcal{G}_{\alpha\beta}(\vec{k},\omega_n) = 
    \frac{\delta_{\alpha\beta}}{i\omega_{n} -
   \left[\epsilon_{\vec{k}}^{(0)}+(-1)^{\alpha+\frac{1}{2}}\mu_{0}B-\mu\right]/
    \hbar}
\end{equation}
and leads to the magnetization
\begin{eqnarray}
\label{settimo}
  \langle\hat{M}\rangle &=&
  \frac{\mu_0}{\Omega}\int{\d}\vec{x}\int\frac{{\d}\vec{k}}{(2\pi)^3}
    \lim_{\eta\to 0}\frac{1}{\beta\hbar}\sum_{n}e^{i\omega_{n}\eta} 
    \nonumber \\
  && \qquad \times \left\{ 
    \frac{1}
    {i\omega_{n}-\left[\epsilon_{\vec{k}}^{(0)}-\mu_{0}B-\mu\right]/\hbar}-
    \frac{1}
    {i\omega_{n}-\left[\epsilon_{\vec{k}}^{(0)}+\mu_{0}B-\mu\right]/\hbar}
    \right\} \nonumber \\
  &=& \mu_0 \int \frac{{\d}\vec{k}}{(2\pi)^3} \left[
    \frac{1}{e^{\beta(\epsilon_{\vec{k}}^{(0)}-\mu_{0}B-\mu)}+1}- 
    \frac{1}{e^{\beta(\epsilon_{\vec{k}}^{(0)}+\mu_{0}B-\mu)}+1}\right] .
\end{eqnarray}
Clearly, Eq.~(\ref{settimo}), together with the condition 
\begin{equation}
  \rho=\frac{\langle\hat{N}\rangle}{\Omega}=\int\frac{{\d}\vec{k}}{(2\pi)^3}
    \left[\frac{1}{e^{\beta(\epsilon_{\vec{k}}^{(0)}-\mu_{0}B-\mu)}+1}+
    \frac{1}{e^{\beta(\epsilon_{\vec{k}}^{(0)}+\mu_{0}B-\mu)}+1}\right]
\end{equation}
fixing the density of the system, in general can be computed only numerically. 
However, if the magnetic field contributes weakly to the energy eigenvalues,
one gets the explicit expression
\begin{eqnarray}
  \langle\hat{M}\rangle &=& -\mu_0\int\frac{{\d}\vec{k}}{(2\pi)^3}
    \left[n(\epsilon_{\vec{k}}^{(0)}+\mu_{0}B)-n(\epsilon_{\vec{k}}^{(0)}
    -\mu_{0}B)\right] \nonumber \\
  &\cong& -2\mu_0^2~B~\int\frac{{\d}\vec{k}}{(2\pi)^3} \left.\left(
    \frac{{\d}n}{{\d}\epsilon_{\vec{k}}}\right)
    \right|_{\epsilon_{\vec{k}}^{(0)}},
\end{eqnarray}
which, as it is well-known, in the $T\to 0$ limit, yields the magnetization
\begin{equation}
  \langle\hat{M}\rangle = \frac{3\mu_0^2\rho B}{2\epsilon_{F}} 
\end{equation}
and the susceptibility
\begin{equation}
  \chi \equiv \frac{\partial \langle\hat{M}\rangle}{\partial B}=
    \frac{3\mu_{0}^{2}\rho}{2\epsilon_{F}},
\label{uffa}
\end{equation}
namely the Pauli paramagnetism. Moreover, in the high temperature limit, the
Curie law 
\begin{equation} 
  \chi=\frac{\mu_0^2\rho}{k_BT}
\end{equation}
is attained.
Thus, Matsubara theory, rather than perturbation theory at $T=0$, yields the
correct result.

The above findings relate to the structure of the thermal propagator, which
sums over all the possible configurations weighted by the statistical operator
and hence leads to the true, symmetry-breaking, ground state in the $T\to 0$
limit. 

In fact, by computing the \textit{real time} Green's function at finite $T$
through a proper analytical continuation of the thermal propagator and then by
taking the vanishing temperature limit, one gets
\begin{equation}
\label{roma}
  G_{\alpha\beta}(\vec{k},\omega)=\delta_{\alpha\beta}
    \left[\frac{\theta(\epsilon_{\vec{k}}^\alpha-\epsilon_F)}
    {\omega-\omega_{k\alpha} + \hbar^{-1}\epsilon_F+i\eta} +
    \frac{\theta(\epsilon_F-\epsilon_{\vec{k}}^\alpha)}
    {\omega-\omega_{k\alpha}+\hbar^{-1}\epsilon_F-i\eta}\right].
\end{equation} 
By comparing the above with (\ref{pino}), we see that an additional Fermi 
energy appears in the denominator and, most importantly, $k_F$ is
replaced by the Fermi energy $\epsilon_F$ and $k$ by
$\epsilon_{\vec{k}}^{\alpha}$. 
Thus, the poles in the complex $\omega$-plane of the propagators (\ref{roma})
and (\ref{pino}) (or the associated spectra) are different: hence the
contributions they provide to the frequency integral no longer cancel out. 

\subsection{Dynamic magnetization: the non-interacting linear response theory}

In this Subsection we briefly derive the magnetization of the free Fermi gas
in the linear response theory at $T=0$. It is well-known that this framework 
leads to the correct result for the Pauli paramagnetism, since it allows a weak
external magnetic field $B(\vec{x},t)$ to break the rotational symmetry of the
non-interacting ground state even at $T=0$, thus leading to magnetization.  

We start by recalling that turning on a perturbation $\hat{H}^{\mathrm{ext}}$ 
at time $t=t_0$, the fluctuation of the vacuum expectation value of a
generic operator $\hat{O}$ at time $t$ reads
\begin{eqnarray}
  \delta<\hat{O}(t)> & \equiv & 
    <\psi_{\mathrm{ext}}(t)|\hat{O}|\psi_{\mathrm{ext}}(t)> -
    <\psi_0|\hat{O}|\psi_{0}> \nonumber \\
  &=& \frac{i}{\hbar}\int_{t_0}^t {\d}t' 
    <\psi_0|\left[\hat{H}_H^{\mathrm{ext}}(t'),\hat{O}_H(t)\right]|\psi_{0}>.
\end{eqnarray}
In our simple example the perturbation is
\begin{equation}
  \hat{H}_H^{\mathrm{ext}}(t) = -\mu_0 \int {\d}\vec{x} 
    \hat{\psi}_H^{\dagger}(\vec{x},t) \sigma^z\hat{\psi}_H(\vec{x},t) 
    B(\vec{x},t)
\end{equation} 
and the role of $\hat{O}$ is played by
\begin{equation}
  \hat{\sigma}_H^z(\vec{x},t)=\hat{\psi}_H^{\dagger}(\vec{x},t)
    \sigma^z\hat{\psi}_H(\vec{x},t) ,
\end{equation}
$|\psi_{0}>$ being the Fermi sphere.

Then, the magnetization along the $z$ axis, namely the magnetic moment of
the volume element, 
\begin{equation}
  <\hat{M}_z(\vec{x},t)> = \mu_0<\hat{\sigma}^z(\vec{x},t)> ,
\end{equation}
is easily expressed as follows
\begin{eqnarray}
\label{milano}
   \hskip -0.8cm <\hat{M}_z(\vec{x},t)> & = & \frac{i}{\hbar}\mu_0^2 \int
     {\d}\vec{x}' {\d}t' \theta(t-t')
    <\phi_0|\left[\hat{\sigma}_H^{z}(\vec{x},t),\hat{\sigma}_H^z(\vec{x}',t')
    \right]| \phi_{0}> \nonumber B(\vec{x}',t') \nonumber \\
  &=& -\mu_0^2 \int {\d}\vec{x}'{\d}t' 
    \Pi^{0R}(\vec{x}t,\vec{x}'t')B(\vec{x}',t'),
\end{eqnarray}
where the \textit{retarded spin-spin polarization propagator} at 
zero-order 
\begin{equation}
\label{firenze}
  i\hbar\Pi^{0R}_{z,z}(\vec{x}t,\vec{x}'t') = \theta(t-t')<\phi_0|
    \left[\hat{\sigma}_H^{z}(\vec{x},t),\hat{\sigma}_H^z(\vec{x}',t')\right]|
    \phi_{0}> 
\end{equation}
has been introduced. By Fourier-transforming (our system is translationally
invariant), Eq.~(\ref{milano}) becomes, for $\omega>0$, 
\begin{equation}
  <\hat{M}_z(\vec{k},\omega)> = -\mu_0^2 \Pi^{0}(\vec{k},\omega)
    B(\vec{k},\omega),
\end{equation}
$\Pi^{0}(\vec{k},\omega)$ being the familiar free Fermi gas polarization
propagator \cite{Fet71}.
The above, in the $\omega\to0^+$ and $k\to0$ limits, yields the magnetization
induced by a static and uniform magnetic field. Since \cite{Fet71}
\begin{equation}
\label{sciacca}
  \lim_{k\to 0}\mathrm{Re}\Pi^{0}(\vec{k},0) = 
    -\frac{mk_F}{\hbar^2\pi^2}=
    -\frac{3}{2}\frac{\rho}{\epsilon_{F}},
\end{equation}
one gets
\begin{equation}
  <\hat{M}_z>=\frac{3}{2}\frac{\rho}{\epsilon_F}\mu_0^2 B,
\end{equation}
thus recovering the correct susceptibility given by equation (\ref{uffa}).

In Eq.~(\ref{sciacca}), the $\omega\to0$ limit, which implies a static external
field, should be taken \textit{before} the $k\to0$ one, which implies a uniform
external field. 

\subsection{Static magnetization: the propagator for the interacting Fermi
  system} 

We now enlarge the previous analysis by switching on the neutron-neutron 
interaction
\begin{equation}
\label{interazionesolosigma}
  V({\vec{x}}_1-{\vec{x}}_2) =\vec{\sigma}_{1} \cdot \vec{\sigma}_{2} 
    V_{1}~\frac{\lambda^{2}}{4\pi}
    \frac{e^{-\lambda |{\vec{x}}_1-{\vec{x}}_2|}}{|{\vec{x}}_1-{\vec{x}}_2|}.
\end{equation}
The first order self-energy splits now into two pieces that, at finite $T$,
read 
\begin{subequations}
\label{sigmapiupiumenomeno}
\begin{eqnarray}
\label{sigmapiupiu}
 \hskip -0.2cm \Sigma^{\star++}_{(1)}(\vec{k},T) &=&
    - \int \frac{{\d} \vec{k}_{1}}{(2\pi)^{3}}
    v(\vec{k}-\vec{k}_{1})
    \left[ \frac{1}{e^{\beta(\epsilon^{(0)}_{\vec{k}_{1}}-\mu_{0}B-\mu)}+1}+
    \frac{2}{e^{\beta(\epsilon^{(0)}_{\vec{k}_{1}}+\mu_{0}B-\mu)}+1}\right]
    \nonumber \\
  && + V_1\int \frac{{\d} \vec{k}_{1}}{(2\pi)^{3}}
    \left[ \frac{1}{e^{\beta(\epsilon^{(0)}_{\vec{k}_{1}}-\mu_{0}B-\mu)}+1}-
    \frac{1}{e^{\beta(\epsilon^{(0)}_{\vec{k}_{1}}+\mu_{0}B-\mu)}+1}\right]
\end{eqnarray}
and
\begin{eqnarray}
\label{sigmamenomeno}
   \hskip -0.2cm \Sigma^{\star--}_{(1)}(\vec{k},T) &=&
   - \int \frac{{\d} \vec{k}_{1}}{(2\pi)^{3}}
     v(\vec{k}-\vec{k}_{1})
     \left[ \frac{2}{e^{\beta(\epsilon^{(0)}_{\vec{k}_{1}}-\mu_{0}B-\mu)}+1}+
     \frac{1}{e^{\beta(\epsilon^{(0)}_{\vec{k}_{1}}+\mu_{0}B-\mu)}+1}\right]
    \nonumber \\
  && - V_1\int \frac{{\d} \vec{k}_{1}}{(2\pi)^{3}}
    \left[ \frac{1}{e^{\beta(\epsilon^{(0)}_{\vec{k}_{1}}-\mu_{0}B-\mu)}+1}-
    \frac{1}{e^{\beta(\epsilon^{(0)}_{\vec{k}_{1}}+\mu_{0}B-\mu)}+1}\right].
\end{eqnarray}
\end{subequations}
The associated HF equations then become
\begin{equation}
\label{sistematrino}
  \left\{\begin{array}{rl}
    \epsilon_{\vec{k}}^{+} &=\epsilon_{\vec{k}}^{(0)}-\mu_{0}B-
    {\displaystyle \int
    \frac{{\d} \vec{k}_{1}}{(2\pi)^{3}} 
    v(\vec{k}-\vec{k}_{1})
    \left( \frac{1}{e^{\beta(\epsilon^{+}_{\vec{k}_{1}}-\mu)}+1}+
    \frac{2}{e^{\beta(\epsilon^{-}_{\vec{k}_{1}}-\mu)}+1}\right) } \\
    & \quad + V_1 {\displaystyle \int \frac{{\d} \vec{k}_{1}}{(2\pi)^{3}}
    \left( \frac{1}{e^{\beta(\epsilon^{+}_{\vec{k}_{1}}-\mu)}+1}-
    \frac{1}{e^{\beta(\epsilon^{-}_{\vec{k}_{1}}-\mu)}+1}\right) } \\
  \epsilon_{\vec{k}}^{-} &=\epsilon_{\vec{k}}^{(0)}+\mu_{0}B-
    {\displaystyle \int \frac{{\d}
    \vec{k}_{1}}{(2\pi)^{3}} 
     v(\vec{k}-\vec{k}_{1})
     \left( \frac{2}{e^{\beta(\epsilon^{+}_{\vec{k}_{1}}-\mu)}+1}+
     \frac{1}{e^{\beta(\epsilon^{-}_{\vec{k}_{1}}-\mu)}+1}\right) } \\
    & \quad - V_1 {\displaystyle \int \frac{{\d} \vec{k}_{1}}{(2\pi)^{3}}
    \left( \frac{1}{e^{\beta(\epsilon^{+}_{\vec{k}_{1}}-\mu)}+1}-
    \frac{1}{e^{\beta(\epsilon^{-}_{\vec{k}_{1}}-\mu)}+1}\right) } \\
  \rho &= {\displaystyle \int \frac{{\d} \vec{k}_{1}}{(2\pi)^{3}}\left(
    \frac{1}{e^{\beta(\epsilon^{+}_{\vec{k}_{1}}-\mu)}+1}+ 
     \frac{1}{e^{\beta(\epsilon^{-}_{\vec{k}_{1}}-\mu)}+1}\right) }.
  \end{array} \right .
\end{equation}
From Eqs.~(\ref{sigmapiupiumenomeno}) and (\ref{sistematrino}) it appears that,
owing to the presence of the magnetic field $B$, the self-energy not only is no
longer proportional to the unit matrix in spin space, but it acquires, {\em 
beyond the exchange contribution, a direct one as well}, an occurrence with far
reaching consequences.

\begin{figure}
\begin{center}
\includegraphics[clip,width=\textwidth]{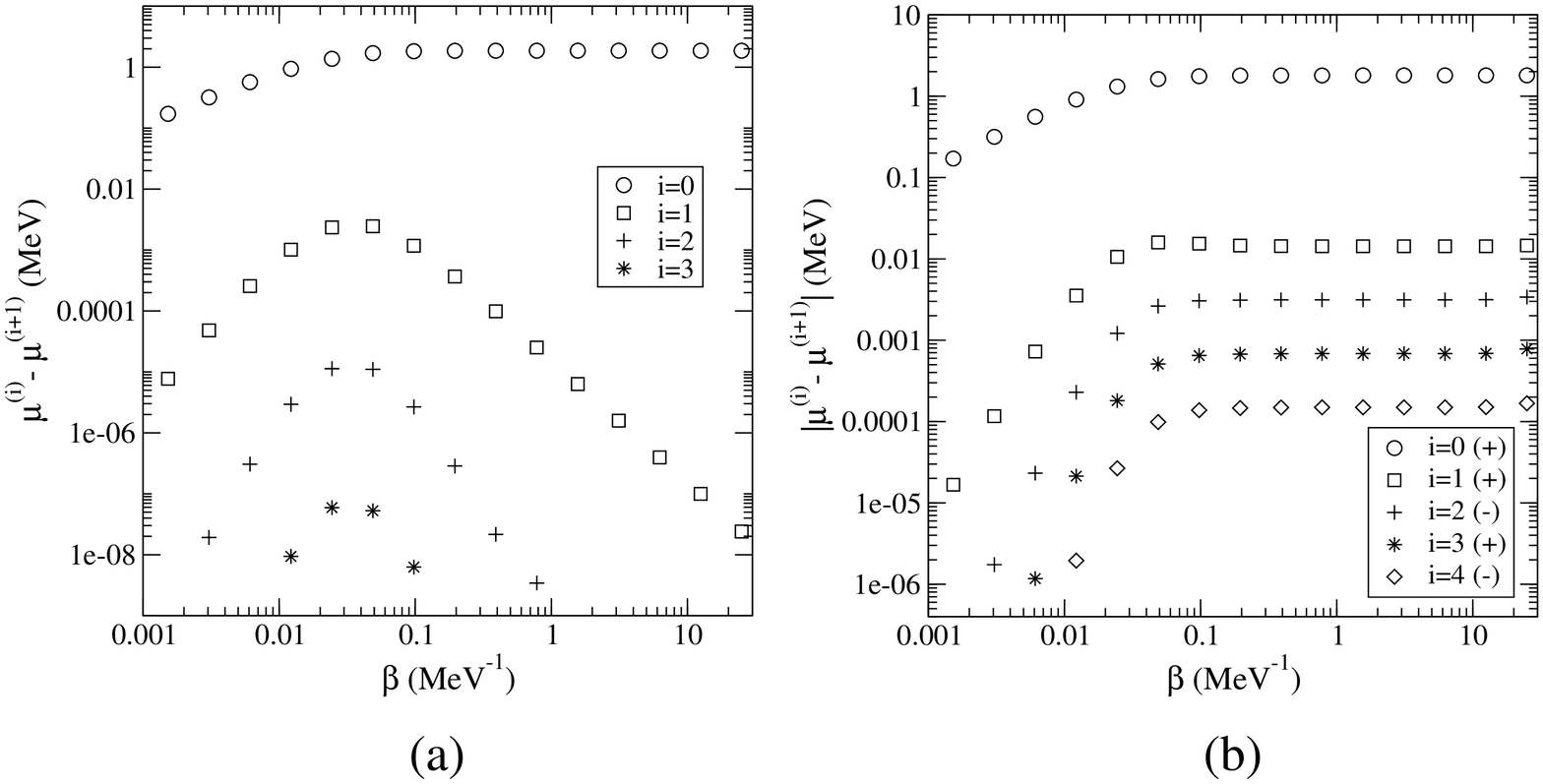}
\caption{\label{fig:pione050differ} The difference between the successive HF
  iterations $\mu^{(i)}-\mu^{(i+1)}$
 as a function of $\beta$ in the case of a finite range interaction with
 $\lambda=140$~MeV and $V_1=50$~MeV~fm$^3$ in absence of magnetic field
 (a) and with a background magnetic field $B=10^{14}$~tesla (b); 
 $\rho=0.17$~fm$^{-3}$. In panel (b) the modulus of the differences is
 displayed: note the alternation in sign that implies an oscillating approach
 to $\mu_{\mathrm{HF}}$. } 
\vskip 1cm
\includegraphics[clip,width=\textwidth]{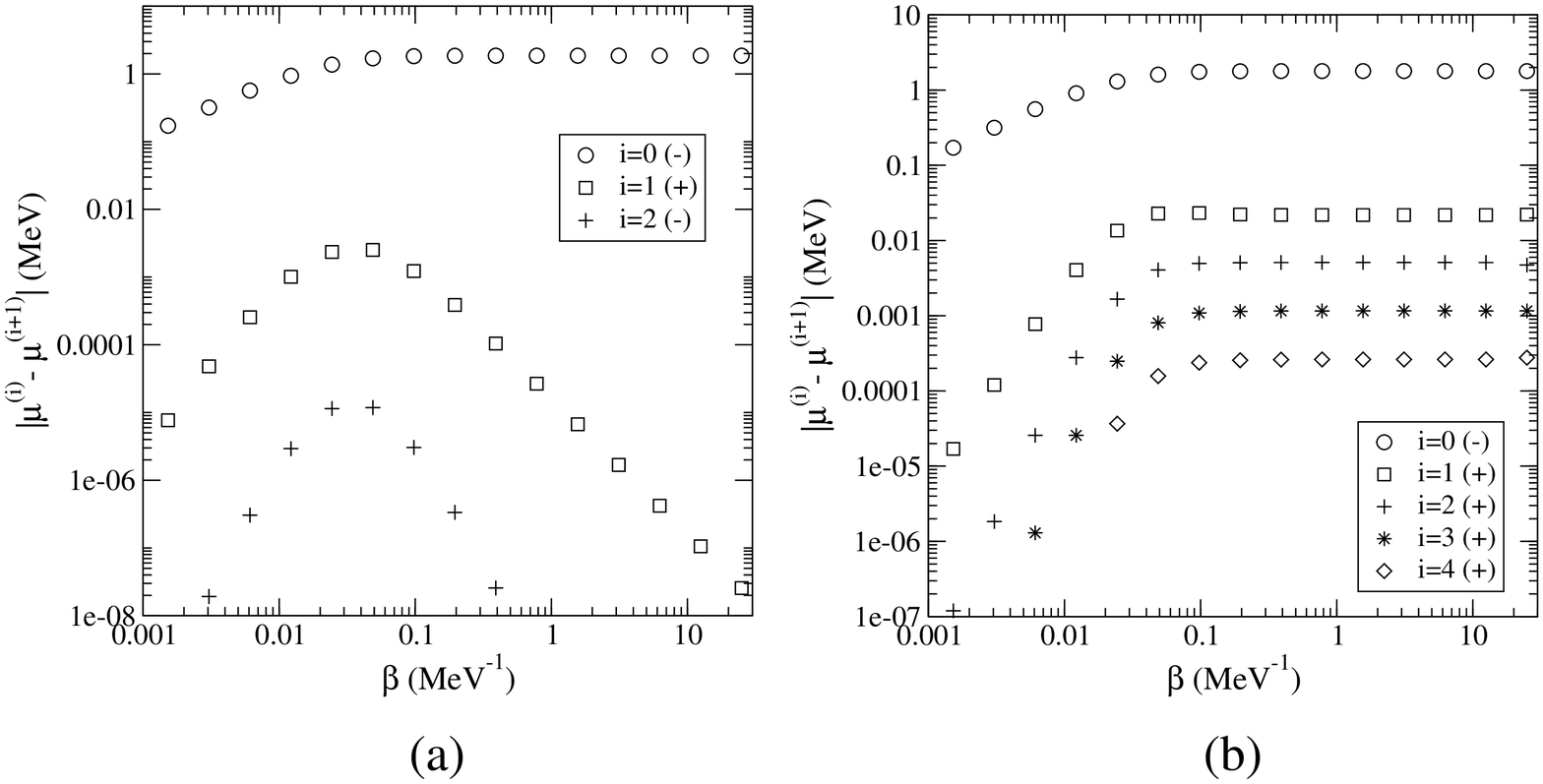}
\caption{\label{fig:pione-050differ} As in
  Fig.~\protect\ref{fig:pione050differ}, but with $V_1=-50$~MeV~fm$^3$.
  Note that the differences are now constant in sign, except the first one:
  hence the approach to $\mu_{\mathrm{HF}}$ is smooth. }
\end{center}
\end{figure}

We have solved the system (\ref{sistematrino}) numerically, searching for the
magnetization in Eq.~(\ref{settimo}) and the HF chemical potential
$\mu_{\mathrm{HF}}$. Our results for $\mu_{\mathrm{HF}}$ are displayed in
Figs.~\ref{fig:pione050differ} and \ref{fig:pione-050differ}.
For purpose of illustration we have chosen the very large value of 
$10^{14}$~tesla for $B$ (on the surface of neutron stars $B$ is estimated to 
range from $10^8$ to $10^{10}$~tesla \cite{Dys71}, whereas in the interior its
upper limit is set at a few times $10^{14}$~tesla \cite{Lai91}).

Remarkably, in the presence of a magnetic field the contributions to
$\mu_{\mathrm{HF}}$ arising from the higher iterations 
(Fig.~\ref{fig:pione050differ}b)
are seen to stay quite constant with the temperature till very small $T$,
unlike the case with no symmetry breaking 
(Fig.~\ref{fig:pione050differ}a), where they
reach a maximum for temperatures close to $T_F$. 

In the proximity of $T=0$, the higher order diagrams contributing to
$\mu_{\mathrm{HF}}$, rather than canceling each other, now contribute more than
they do at high $T$.
This is a beautiful manifestation of the failure of the cancellation theorem
in the presence of a symmetry breaking.

Of significance is also the alternating behavior of the iterations for an
antiferromagnetic interaction: 
those of odd order have a positive sign, those of even order a negative sign 
(note that in Fig.~\ref{fig:pione050differ}b we display the
absolute value of $\mu^{(i)}-\mu^{(i+1)}$). As a consequence the approach to
the HF self-consistent solution is no longer smooth, as in the symmetric
case,  but now the successive iterations oscillate around the HF solution
until they stabilize at the latter value when self-consistency is reached. 

In contrast, it is remarkable that for the ferromagnetic coupling the sign of
the successive iterations is constant (positive), except for the first one,
which is negative. Hence, in this case the self-consistent solution is smoothly
reached. The situation is reminiscent of the physics of
an hydrogen atom in an external magnetic field, the competition between the
Coulomb and the magnetic field now being replaced by the one between the
nucleon-nucleon interaction and the magnetic field. If this opposes the action
of the nucleon-nucleon forces, then it makes it harder to reach
self-consistency.

\subsection{Dynamic magnetization: the interacting linear response theory}

As previously seen for the case of a non-interacting Fermi system, also when
the interaction is switched on we can obtain the system's magnetization and 
susceptibility in the framework of the linear response theory at $T=0$ using
the polarization propagator, a tool set up with the two-fermion Green's
function. 

We start by computing the spin-spin retarded polarization propagator needed
in the HF theory. In the present case, owing to the nature of the interaction
in Eq.~(\ref{interazionesolosigma}), only the Fock term is non-vanishing, so we
shall actually compute $\Pi^{\mathrm{F}}$. At $T=0$ its real part reads
\begin{equation}
\label{repihf}
  \mathrm{Re}\Pi^{\mathrm{F}}(\vec{q},\omega) = \frac{2}{\hbar} \mathcal{P} 
    \int \frac{{\d}\vec{k}}{(2\pi)^{3}} 
    \theta(\mid\vec{k}+\vec{q}\mid-k_{F})]\theta(k_{F}-k)
    \frac{2\omega_{\vec{k}\vec{q}}}{\omega^2-(\omega_{\vec{k}\vec{q}})^2},
\end{equation}
where
\begin{equation}
  \omega_{\vec{k}\vec{q}} = \omega_{\vec{k}+\vec{q}} + 
    \Sigma^{\star}_{(1)}(\vec{k}+\vec{q}) -
    \omega_{\vec{k}}-\Sigma^{\star}_{(1)}(\vec{k}).
\end{equation}
The above expression cannot be computed analytically. Yet, 
in the case of the interaction in Eq.~(\ref{interazionesolosigma}), a quite
accurate analytic approximation to Eq.~(\ref{repihf}) can be obtained by
expanding the self-energy in Eq.~(\ref{unofinitozero}) (divided by 3, since we
dropped the isospin degree of freedom) up to and including terms in $k^2$; one
gets 
\begin{equation}
  \Sigma^{\star}_{(1)}(\vec{k}) = A+Bk^2,
\end{equation}
with
\begin{equation}
  A=\frac{3\lambda^3V_{1}}{2\hbar\pi^2} \left(\arctan\frac{k_{F}}{\lambda} -
    \frac{k_F}{\lambda}\right) 
\end{equation}
and
\begin{equation}
  B = \frac{k_{F}^{3}\lambda^{2}V_{1}}{2\hbar\pi^2(k_{F}^{2}+\lambda^{2})^{2}}.
\end{equation}
Then, computing $\Sigma^{\star}(\vec{k}+\vec{q})-\Sigma^{\star}(\vec{k})$ in
this approximation and inserting this result into Eq.~(\ref{repihf}), we get
for $\mathrm{Re}\Pi^{\mathrm{F}}$ an expression identical to the free one, but
for the replacement of the nucleon mass $m$ with an effective mass $m^{\star}$
given by 
\begin{equation}
\label{eq:mstar}
  \frac{m^{\star}}{m}=\left[1+ 
    \frac{k_{F}^{3}\lambda^2mV_{1}}{\hbar^2\pi^2(k_{F}^{2}+\lambda^2)^2}
    \right]^{-1}.
\end{equation}
Next, letting as before first $\omega\to0$ and then $q\to0$, we obtain for the
magnetization and the magnetic susceptibility the expressions
\begin{subequations}
\begin{equation}
\label{magnstar}
  <\hat{M}_z>=\frac{3}{2}\frac{\rho}{\epsilon_F^{\star}}\mu_0^2 B
\end{equation}
and
\begin{equation}
  \chi=\frac{3}{2}\frac{\rho}{\epsilon_F^{\star}}\mu_0^2, 
\end{equation}
\end{subequations}
respectively, where the Fermi energy now reads
$\epsilon_{F}^{\star}=\hbar^2k_F^2/2m^{\star}$. Since, for the
interaction in Eq.~(\ref{interazionesolosigma}) with $V_1>0$, one has
$m^{\star}<m$, we conclude that for an antiferromagnetic force the HF mean
field \textit{lowers} the free magnetization (or the susceptibility). 

However it is of significance that Eq.~(\ref{magnstar}) \textit{does not
coincide} with the $T\to0$ limit of the magnetization 
\begin{equation}
\label{piumeno}
  \langle\hat{M}\rangle = \mu_0 \int\frac{{\d}\vec{k}}{(2\pi)^3}
    \left[ \frac{1}{e^{\beta(\epsilon^+_{\vec{k}}-\mu)}+1} -
    \frac{1}{e^{\beta(\epsilon^-_{\vec{k}}-\mu)}+1} \right],
\end{equation}
$\epsilon^+_{\vec{k}}$, $\epsilon^-_{\vec{k}}$ and $\mu$ being the HF
single-particle energies and chemical potential obtained by solving 
Eq.~(\ref{sistematrino}). To achieve the accord between Eqs.~(\ref{magnstar})
and (\ref{piumeno}) it is necessary to go beyond HF. This will be done in the 
next Section.

\begin{figure}
\begin{center}
\includegraphics[clip,height=8cm]{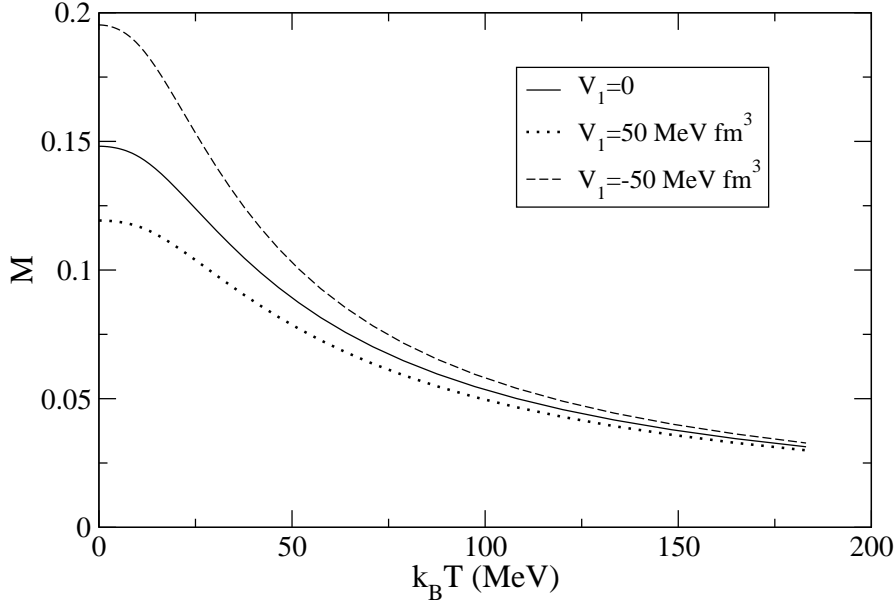}
\caption{\label{fig:magnvst} The HF magnetization in an external magnetic field
 as a function of temperature for $B=10^{14}$~tesla and for different values of
 $V_1$; $\lambda=140$~MeV and $\rho=0.17$~fm$^{-3}$. } 
\end{center}
\end{figure}

Here we display in Fig.~\ref{fig:magnvst} the temperature behavior of the
relative magnetization $M$ (yielding the fraction of spins anti-aligned 
to the magnetic field, i.e.
$M=<\hat{M}_z>/|\mu_0|\rho=(N^\uparrow-N^\downarrow)/(N^\uparrow+N^\downarrow)$
with obvious meaning of the symbols) associated with Eq.~(\ref{piumeno}). Note
the decreasing of $M$ as $T$ increases and the recovery at large $T$ of the
Curie value, the interaction among neutrons becoming irrelevant at high
$T$. Moreover the values of $M$ at $T=0$ are lower (larger) than the free one
for an antiferromagnetic (ferromagnetic) force: they will be shown to coincide
with the predictions of the linear response theory in a frame extending the HF
one.

\section{Beyond HF}

In the framework of the linear response theory the natural extension of the HF
approach should embody the RPA (random phase approximation) correlations. 
To start with, we neglect the exchange terms, i.e. we make use of the simpler 
ring approximation. 

This yields for the spin-spin polarization propagator of a homogeneous infinite
system the expression 
\begin{equation}
\label{eq:Piring}
  \Pi^{\mathrm{ring}}(\vec{q},\omega)=\frac{\Pi^{\mathrm{F}}(\vec{q},\omega)}
    {1-V_1 \frac{\displaystyle
    \lambda^2}{\displaystyle q^2+\lambda^2}\Pi^{\mathrm{F}}(\vec{q},\omega)} ,
\end{equation}
from where the magnetization 
\begin{equation}
\label{mring}
  <\hat{M}>=-\lim_{q\to0}\lim_{\omega\to0}
    \mathrm{Re}\Pi^{\mathrm{ring}}(\vec{q},\omega)\mu_0^2B = 
    \frac{3\rho/2\epsilon_F^{\star}}{1+V_1(3\rho/2\epsilon_F^{\star})}\mu_0^2B,
\end{equation}
which is linear in $B$, is deduced. Thus, for the magnetic susceptibility in
ring approximation, one  
gets 
\begin{equation}
\label{chiring}
  \chi^{\mathrm{ring}} =
  \frac{3\rho/2\epsilon_F^{\star}}{1+V_1(3\rho/2\epsilon_F^{\star})}\mu_0^2 ,
\end{equation}
showing that the ring correlations, for an anti-ferromagnetic $(V_1>0)$
nucleon-nucleon interaction, lower the susceptibility of the free Fermi gas, 
thus acting in the same direction of the HF mean field. 

The quenching of the free Fermi gas susceptibility $\chi^{\mathrm{free}}$ in
ring approximation relates to the existence of a spin collective mode, called a
\textit{magnon}, which makes harder the excitation of the system. As it is
well-known, the energy of the magnon is found by solving the equation
\begin{equation}
\label{eq:magnons}
  1-V_1\frac{\lambda^2}{q^2+\lambda^2}\mathrm{Re}
    \Pi^{\mathrm{F}}(\vec{q},\omega) = 0,
\end{equation}
that --- in the limit $\omega\to0$, $q\to0$, but keeping now the ratio
$x=(\omega/q)m^{\star}/\hbar k_F$ constant and greater than one ---
is fulfilled by the phonon-like dispersion relation $\omega=c_Sq$, the 
zero-sound velocity $c_S$ being fixed by the equation \cite{Fet71}
\begin{equation}
  \Phi(x)=\frac{\pi^2\hbar^2}{m^{\star}k_FV(0)},
\end{equation}
with
\begin{equation}
  \Phi(x)=\frac{1}{2}x\ln\frac{x+1}{x-1}-1.
\end{equation}

To assess the role of anti-symmetrization one resorts to the full RPA,
which includes the exchange diagrams beyond the ring ones. 
This we do with the method of the continued fractions \cite{Len80,Fes92,DeP98},
which, when truncated at the first fraction, yields for the RPA polarization
propagator the following expression 
\begin{equation}
\label{eq:PiRPA}
  \Pi^{\mathrm{RPA}}(\vec{q},\omega)=\frac{\Pi^{\mathrm{F}}(\vec{q},\omega)}
    {1-V_1 \frac{\displaystyle
    \lambda^2}{\displaystyle q^2+\lambda^2}\Pi^{\mathrm{F}}(\vec{q},\omega)
    -\Pi^{\mathrm{F}(1)}_{\text{ex}}(\vec{q},\omega) /
    \Pi^{\mathrm{F}}(\vec{q},\omega)}, 
\end{equation}
where 
\begin{equation}
  \Pi^{\mathrm{F}(1)}_{\text{ex}}(\vec{q},\omega) = -
    \int\frac{{\d}\vec{k}_1}{(2\pi)^3} 
    E(\vec{k}_1,\vec{q}) \int\frac{{\d}\vec{k}_2}{(2\pi)^3}
    E(\vec{k}_2,\vec{q}) v(\vec{k}_1-\vec{k}_2),
\end{equation}
being
\begin{equation}
  E(\vec{k},\vec{q}) =\frac{1}{\hbar}\left[
    \frac{\theta(|\vec{k}+\vec{q}|-k_F)\theta(k_F-k)} 
    {\omega-\omega_{\vec{k}+\vec{q}}+\omega_{\vec{k}} + i \eta} -
    \frac{\theta(k_F-|\vec{k}+\vec{q}|)\theta(k-k_F)}
    {\omega-\omega_{\vec{k}+\vec{q}}+\omega_{\vec{k}} - i \eta}\right],
\end{equation}
is the exchange diagram associated to the first order ring contribution.
The expression in Eq.~(\ref{eq:PiRPA}), while exact for a zero-range force,
remains a good approximation for typical nuclear finite range interactions too.

One can verify that the magnetization deduced from Eq.~(\ref{eq:PiRPA})
embodies a susceptibility very close to the exact RPA one, since the RPA
problem can be \textit{analytically} solved for an infinite system in the limit
of vanishing frequencies and momenta, where it merges with the Landau's 
quasi-particle theory \cite{Alb82a,Alb82b}. 
One gets (see Appendix~\ref{app:exactRPA})
\begin{equation}
\label{eq:chiRPA}
  \chi^{\mathrm{RPA}} = \frac{3\rho/2\epsilon_F^{\star}}
    {1+V_1(3\rho/2\epsilon_F^{\star})+
    \half V_1 (3\rho/2\epsilon_F^{\star}) (\lambda/2k_F)^2
    \ln\left[1+(2k_F/\lambda)^2\right] }\mu_0^2 ,
\end{equation}
which, for a zero-range force, becomes
\begin{equation}
\label{eq:chiRPAzerorange}
  \chi^{\mathrm{RPA}}_{\mathrm{0-range}} = \frac{3\rho/2\epsilon_F}
    {1+9V_1\rho/4\epsilon_F},
\end{equation}
an expression identical to the one deduced from (\ref{eq:PiRPA}).
Eq.~(\ref{eq:chiRPA}) shows that, in the anti-ferromagnetic case, the
anti-symmetrization reinforces (mildly) the quenching of $\chi^{\mathrm{free}}$
induced by the ring correlations. 

Note that in RPA the impact of the exchange diagrams decreases as the range
of the force increases, as it should: this result follows by comparing
Eq.~(\ref{eq:chiRPA}) with Eq.~(\ref{eq:chiRPAzerorange}). 

\begin{figure}
\begin{center}
\includegraphics[clip,width=8cm]{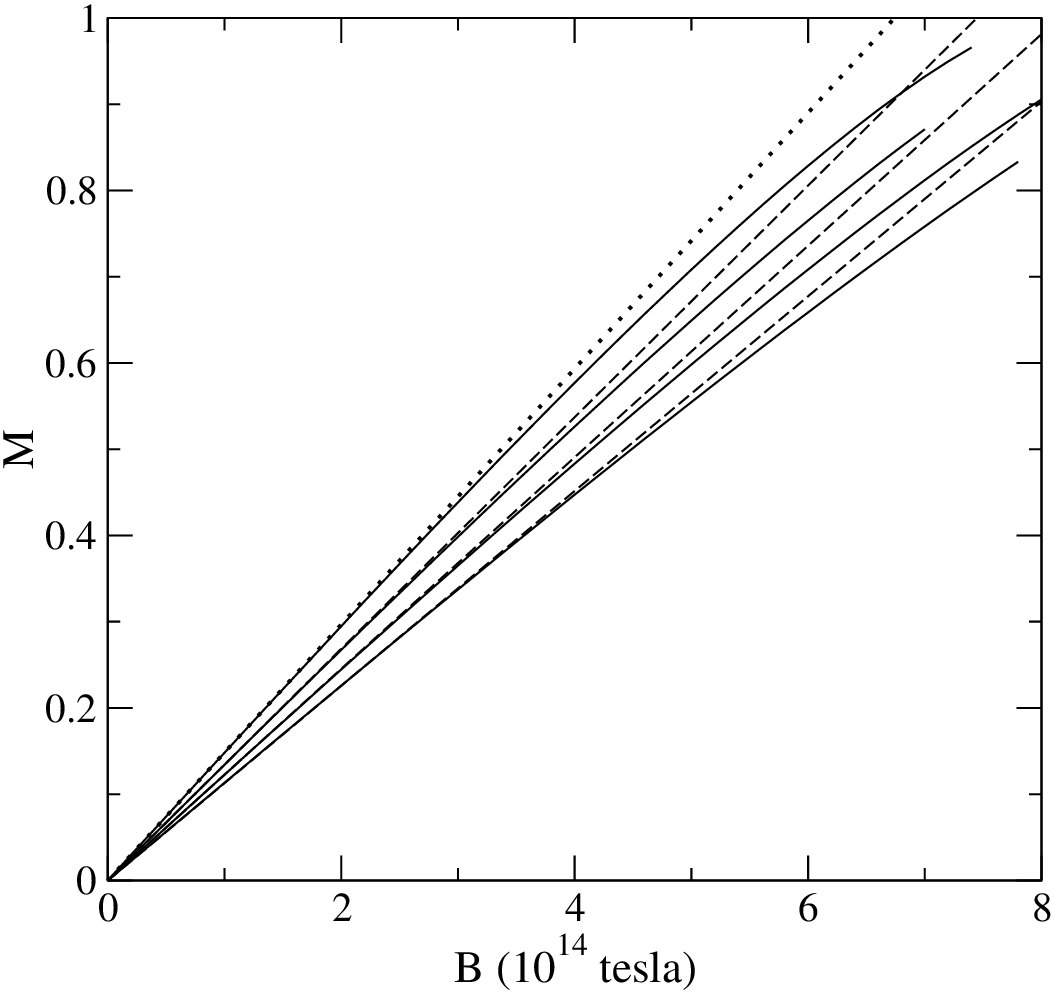}
\caption{\label{fig:mvsbv50_zr} The magnetization as a function of $B$ for 
 $V_1=50$~MeV~fm$^3$ and zero range at $T=0$ ($\rho=0.17$~fm$^{-3}$).
 The dotted lines display the results in the linear response framework (using
 $\Pi$), the solid lines in the thermal self-consistent calculations at $T=0$
 (using $\mathcal{G}^B$). 
 Starting from above the pairs of curves correspond, respectively, to the
 following cases: $\Pi^0$/$\mathcal{G}_0^B$, 
 $\Pi^{\text{ladder}}$/$\mathcal{G}_\text{F}^B$, 
 $\Pi^{\text{ring}}$/$\mathcal{G}_\text{H}^B$,
 $\Pi^{\text{RPA}}$/$\mathcal{G}_{\text{HF}}^B$. }
\vskip 1cm
\includegraphics[clip,width=8cm]{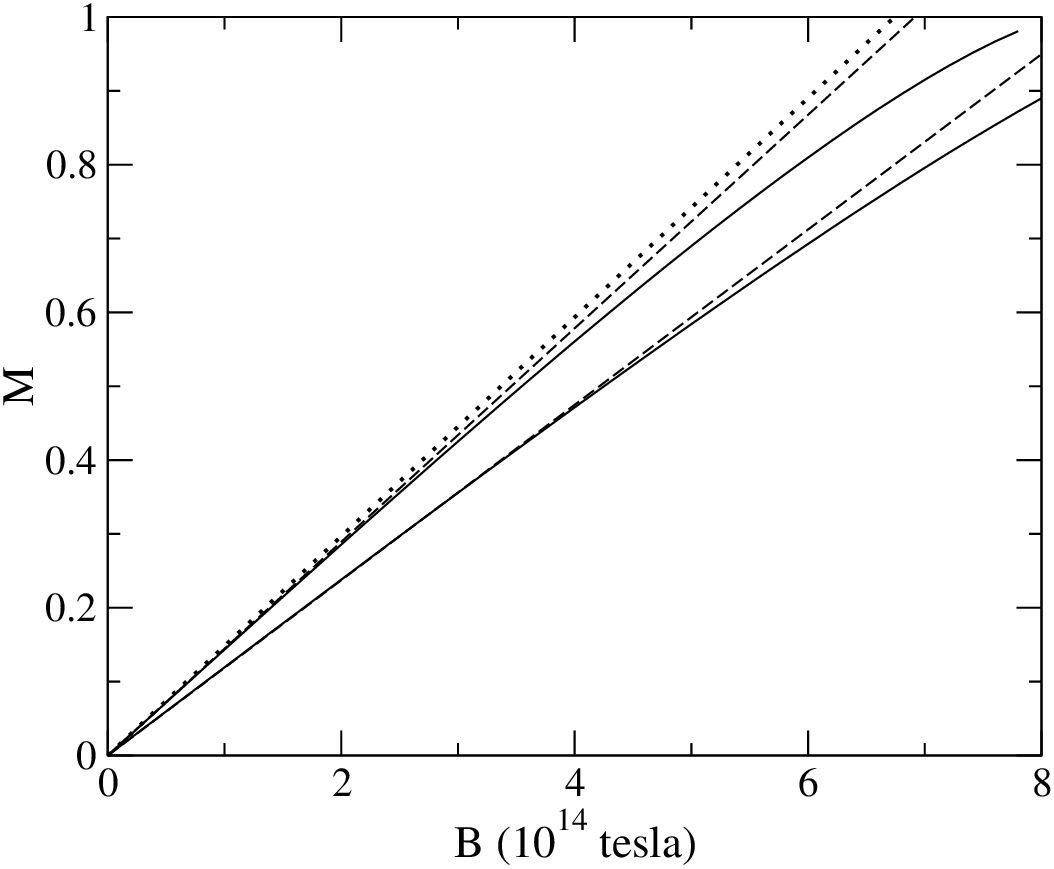}
\caption{\label{fig:mvsbv50} The magnetization as a function of $B$ for 
 $V_1=50$~MeV~fm$^3$ and $\lambda=140$~MeV at $T=0$ ($\rho=0.17$~fm$^{-3}$). 
 The dotted line displays the result in the free linear response framework
 ($\Pi^0$), the upper dashed line in the linear response framework with Fock
 correlations ($\Pi_\text{F}$) and the lower dashed line with Fock and RPA
 correlations ($\Pi_\text{F}^{\text{RPA}}$); the upper solid line represents
 the results of the thermal self-consistent calculations at $T=0$ including
 only Fock correlations ($\mathcal{G}_\text{F}^B$), the lower solid line
 includes HF correlations ($\mathcal{G}_\text{HF}^B$). }
\end{center}
\end{figure}
We are now in a position to find out the polarization propagator yielding (as
long as the response of the system is linear) the HF Matsubara
magnetization. For the sake of simplicity we start by considering a zero-range
antiferromagnetic force of strength $V_1=50$ MeV. In
Figs.~\ref{fig:mvsbv50_zr} and \ref{fig:mvsbv50}
are displayed versus $B$ various relative magnetizations as obtained from the
Matsubara propagator and from the zero-temperature linear response theory. 
In Fig.~\ref{fig:mvsbv50_zr} we employ a zero-range force, hence the HF mean
field does not contribute to the polarization propagators; in
Fig.~\ref{fig:mvsbv50} a finite range force. The figures clearly show that the
magnetization obtained from $\Pi^{RPA}$ coincides with the one given by the HF
thermal propagator $\mathcal{G}_{HF}^B$ for $T\to0$ until $\bar{B}\simeq 3$
(in units of $10^{14}$~tesla), which sets the limit of validity of the linear
response framework. For $B$ larger than $\bar{B}$ the magnetization increases
much less rapidly than linearly until it saturates when 
the system is fully magnetized, an expected behavior correctly reproduced by
the thermal HF theory. We thus conclude that in order to obtain the
magnetization and, in general, any mean value using the single-particle thermal
propagator $\mathcal{G}_{HF}^B$, it is necessary to sum up an infinite series
of loop diagrams in the particle-hole polarization propagator, namely those of 
RPA. This correspondence extends also to the pure ring and ladder polarization
propagator as illustrated in Table~\ref{tabellina} (with $\Pi^{\text{ladder}}$
we mean the propagator summing up only the particle-hole exchange diagram).
\begin{table}[t]
\begin{center}  
\begin{tabular}{|l|l|l|}
\hline
Susceptibility & Feynman, $T=0$ & Matsubara, $T\to 0$ \\
\hline
$\chi^{\text{free}}$ & $\Pi^0$ & $\mathcal{G}_0^B$ \\
\hline
$\chi^{\text{ring}}$ & $\Pi^{\text{ring}}$ & $\mathcal{G}_\text{H}^B$ \\
\hline
$\chi^{\text{ladder}}$ & $\Pi^{\text{ladder}}$ & $\mathcal{G}_\text{F}^B$ \\
\hline
$\chi^{\text{RPA}}$ & $\Pi^{\text{RPA}}$ & $\mathcal{G}_{\text{HF}}^B$ \\
\hline
\end{tabular}
\caption{\label{tabellina} The zero-temperature polarization propagator
  required to obtain the same magnetic susceptibility deduced (in the $T\to 0$
  limit) from the Matsubara free ($\mathcal{G}_0^B$), Hartree
  ($\mathcal{G}_\text{H}^B$), Fock ($\mathcal{G}_\text{F}^B$) and HF 
  ($\mathcal{G}_{\text{HF}}^B$) thermal single-particle propagator. 
  Clearly the many body diagrams summed up to infinite order to get the
  appropriate $\Pi$ are the ring, the ladder and the RPA ones, respectively.} 
\end{center} 
\end{table}

Also worth noticing is that the range of validity of the linear response theory
appears to be only moderately affected  by the many body scheme employed.

Turning now to comment on the finite range force, we observe that the slopes of
the curves in Fig.~(\ref{fig:mvsbv50}) are somewhat reduced with respect to
those in Fig.~(\ref{fig:mvsbv50_zr}).  
Indeed, in the former the fermionic lines in the polarization propagator are
dressed by a $k^2$ dependent Fock self-energy and this, as previously shown,
reduces the susceptibility for an antiferromagnetic interaction. Furthermore,
as expected, $\Pi_\text{F}^{\text{ring}}$ and $\Pi_\text{F}^{\text{RPA}}$ tend 
to become very similar as the range of the force increases. 
  
Thus we have reached the result that the statistical  average
required in computing $\mathcal{G}_{\text{HF}}^B$ is performed  over the states
entering into the spectral representation of the RPA polarization propagator
with fermionic lines dressed by a Fock self-energy.

\begin{figure}
\begin{center}
\includegraphics[clip,height=7cm]{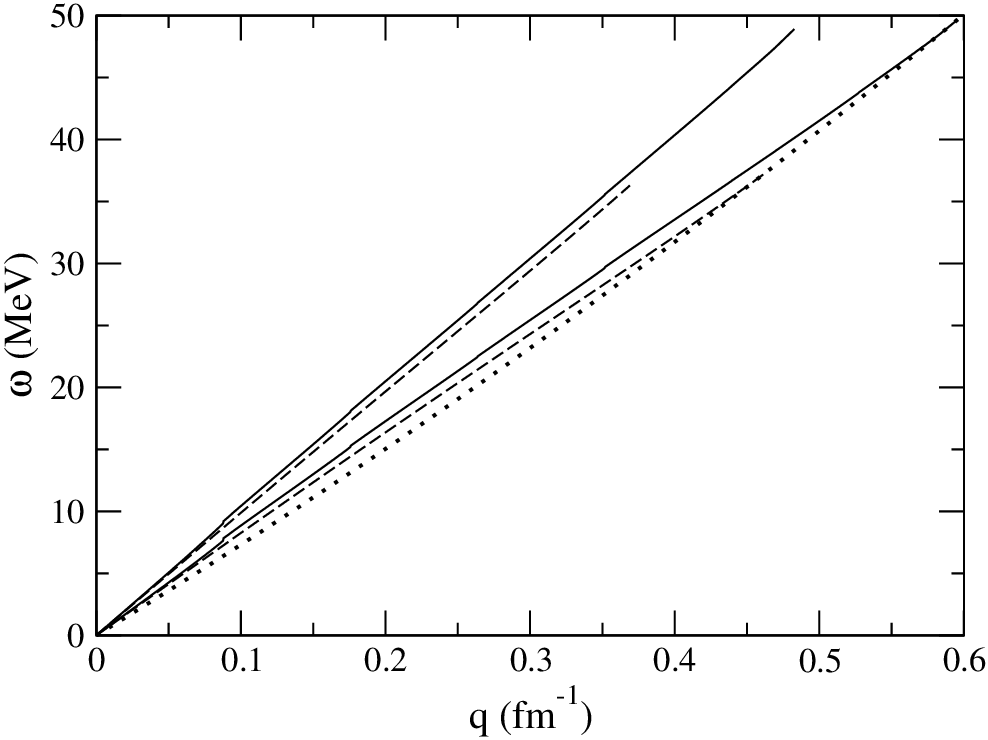}
\caption{\label{fig:magnons} The dispersion relation for the collective modes
  (magnons) in ring (dashed) and RPA (solid), with (upper lines) and without
  (lower lines) the HF mean field for an antiferromagnetic interaction with
  $V_1=480$~MeV~fm$^3$ and $\lambda=140$~MeV ($\rho=0.17$~fm$^{-3}$); the
  dotted line represents the upper border of the ph response region,
  $q^2/2m+qk_F/m$. } 
\vfill
\includegraphics[clip,width=\textwidth]{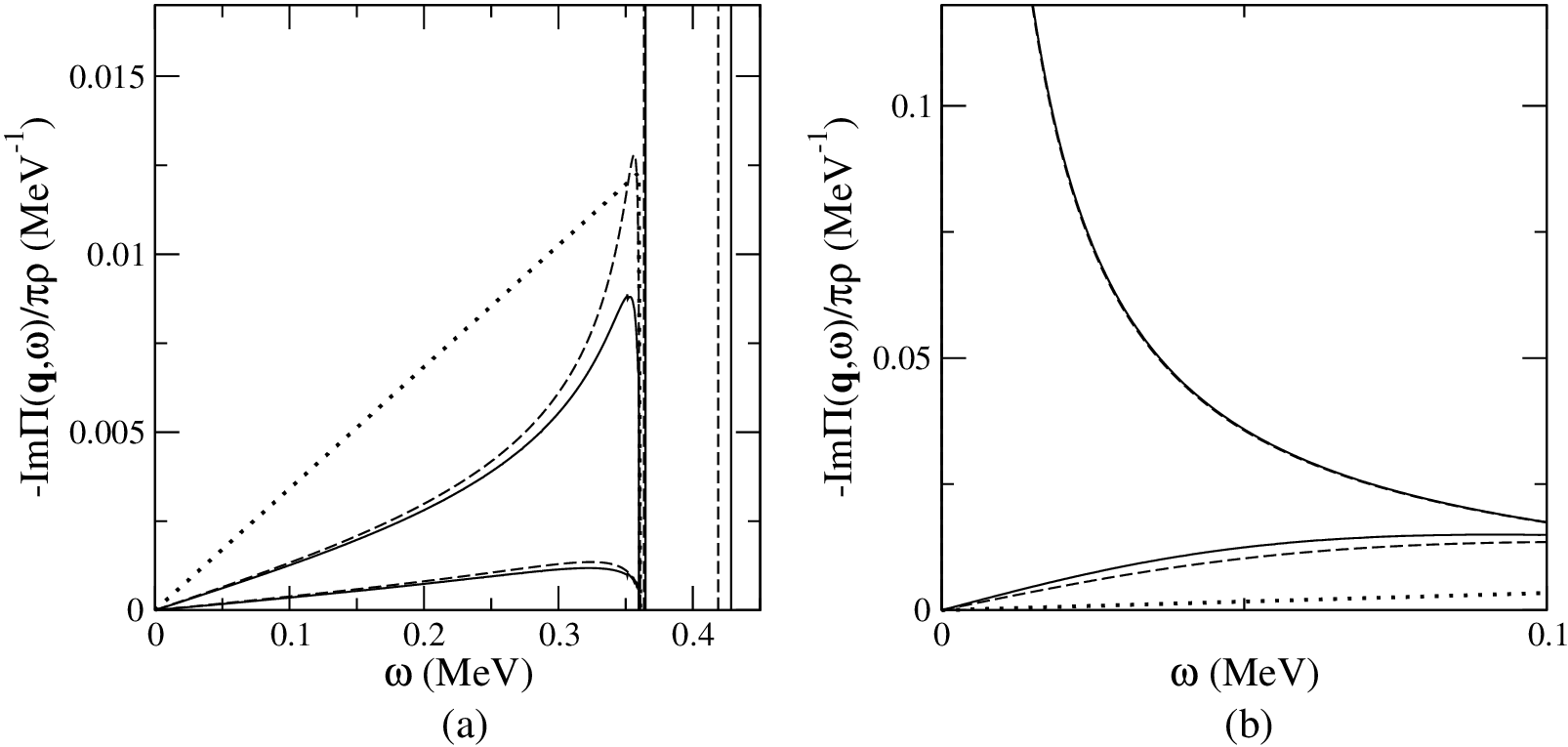}
\caption{\label{fig:respRPA} Spin response functions at $q=1$~MeV/c,
  $\rho=0.17$~fm$^{-3}$ and for a finite range interaction with
  $\lambda=140$~MeV. The dotted lines represent the free response, the dashed
  and solid lines the ring and RPA ones, respectively.
  In panel (a) we employ antiferromagnetic couplings: $V_1=150$~MeV~fm$^3$
  (upper curves) and $V_1=480$~MeV~fm$^3$ lower curves; the spikes represent
  the ring and RPA collective states: the pair at higher energy corresponds to
  $V_1=480$~MeV~fm$^3$, the other one to $V_1=150$~MeV~fm$^3$ (note that in the
  latter case the ring and RPA excitations energies are very close).
  In panel (b) we employ ferromagnetic couplings: $V_1=-150$~MeV~fm$^3$ (lower
  curves), $V_1=V_{1,\text{crit}}^{\text{ring}}\protect\cong-239$~MeV~fm$^3$
  and $V_1=V_{1,\text{crit}}^{\text{RPA}}\protect\cong-223$~MeV~fm$^3$ (note
  that the ring response at $V_1=V_{1,\text{crit}}^{\text{ring}}$ and the RPA
  one at $V_1=V_{1,\text{crit}}^{\text{RPA}}$ practically overlap each other).
  For the sake of simplicity, the HF mean field has not been included. }
\end{center}
\end{figure}
In concluding this Section we display in Fig.~\ref{fig:magnons} the frequency
behavior of the magnons, both in ring and in RPA, with and without the HF mean
field, for an interaction with $V_1=480$~MeV~fm$^3$ and $\lambda=140$~MeV.
The hardening of the magnon mode stemming from the HF mean field and (to less
extent) from anti-symmetrizing the ring diagrams is clearly apparent in the
figure. 

We display also in Fig.~\ref{fig:respRPA} the system's spin response functions
(proportional to the imaginary part of $\Pi(\vec{q},\omega)$) both in ring and
in RPA and both for an antiferromagnetic and a ferromagnetic coupling at
$q=1$~MeV/c. 
For sake of illustrating the impact of the pure ring and RPA correlations on
the response we ignore in the figure the action of the Fock mean field (namely,
we compute Eqs.~(\ref{eq:Piring}) and (\ref{eq:PiRPA}) replacing $\Pi^\text{F}$
with $\Pi^0$, or, in other words, we set $m^*=m$).
The values chosen for the coupling are $V_1=\pm150, 480, -239$ and
$-223$~MeV~fm$^3$, whereas the range parameter is $\lambda=140$~MeV. 
In the figure it is clearly seen that: 
\begin{itemize}
\item[a)] for an anti-ferromagnetic coupling, as expected, the magnons are
  standing out above the particle-hole (ph) continuum, the more so the larger
  $V_1$ is. Correspondingly, the depletion of the ph continuum increases with
  $V_1$. Note also that the impact of anti-symmetrization is modest, as
  expected owing to the long range of the force: the more  so, the larger $V_1$
  is; 
\item[b)] for the ferromagnetic coupling the response is of course enhanced,
  the enhancement becoming however dramatic for $V_1=-239$~MeV~fm$^3$ in ring 
  and for $V_1=-223$~MeV~fm$^3$ in RPA. These values of the coupling are
  those yielding a divergent ring and RPA susceptibility, respectively, when 
  $m^*=m$. We shall return on this point in the next Section. 
\end{itemize}

\section{The spontaneous symmetry breaking}

From the explicit formulas of $\chi^{\mathrm{ring}}$ and $\chi^{\mathrm{RPA}}$
obtained in the linear response scheme one sees that both susceptibilities
diverge in the case of a ferromagnetic coupling in correspondence to the
following critical values of the strength of the force:
\begin{subequations}
\label{eq:V1crit}
\begin{equation}
\label{eq:V1critring}
  V_{1,\mathrm{crit}}^{\mathrm{ring}} = - \frac{2\epsilon^*_F}{3\rho}
   \Rightarrow  
   V_{1,\mathrm{crit}}^{\mathrm{ring}} = - \frac{\pi^2\hbar^2}{mk_F} 
   \frac{(1+\bar\lambda^2)^2}{\bar\lambda^4+3\bar\lambda^2+1}
\end{equation}
and
\begin{eqnarray}
\label{eq:V1critRPA}
  V_{1,\mathrm{crit}}^{\mathrm{RPA}} &=& - \frac{2\epsilon^*_F}{3\rho}
    \frac{1}{1+\half(\lambda/2k_F)^2\ln\left[1+(2k_F/\lambda)^2\right]}
    \Rightarrow \nonumber \\ 
  V_{1,\mathrm{crit}}^{\mathrm{RPA}} &=& - \frac{\pi^2\hbar^2}{mk_F} 
    \frac{(1+\bar\lambda^2)^2}{\bar\lambda^4+3\bar\lambda^2+1 + 
    \frac{1}{2}(1+\bar\lambda^2)^2 (\bar\lambda/2)^2 \ln [1+(2/\bar\lambda)^2]}
\end{eqnarray}
\end{subequations}
with $\bar\lambda=\lambda/k_F$. The above expressions have been deduced making
use of Eq.~(\ref{eq:mstar}), which yields the neutron effective mass $m^*$ in
terms of $V_1$, $\lambda$ and $k_F$.

\begin{figure}
\begin{center}
\includegraphics[clip,height=8cm]{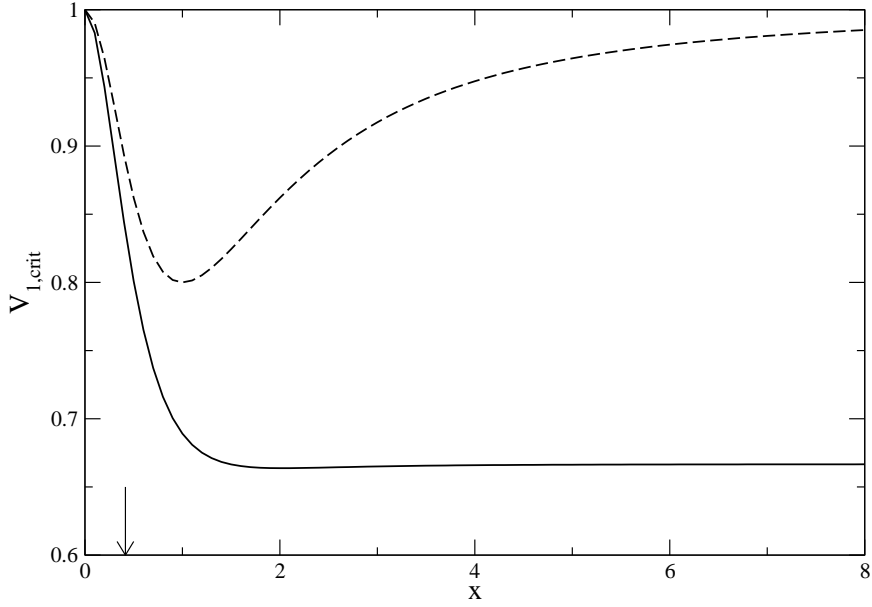}    
\caption{\label{fig:vcritico}The behavior of the critical interactions
  $V_{1,\mathrm{crit}}^{\mathrm{ring}}$ (dashed) and
  $V_{1,\mathrm{crit}}^{\mathrm{RPA}}$ (solid) versus $\bar\lambda=\lambda/k_F$
  in units of $-\pi^2\hbar^2/mk_F$. The arrow corresponds to $\lambda=140$~MeV
  and $k_F=1.71$~fm$^{-1}$. }
\end{center}
\end{figure}

The behavior of Eqs.~(\ref{eq:V1critring}) and (\ref{eq:V1critRPA}) versus
$\bar\lambda=\lambda/k_F$ is displayed in Fig.~\ref{fig:vcritico}: note that
the two curves coincide for an infinite range force ($\bar\lambda=0$) whereas
their difference is the largest for a contact interaction
($\bar\lambda=\infty$). For $\lambda=140$ MeV and $k_F=1.71$~fm$^{-1}$ one has 
$V_{1,\mathrm{crit}}^{\mathrm{ring}}\cong-239$~MeV~fm$^3$ and 
$V_{1,\mathrm{crit}}^{\mathrm{RPA}}\cong-223$~MeV~fm$^3$: 
these are the values previously employed (see Fig.~\ref{fig:respRPA}). 

In this Section, for sake of simplicity, we shall consider a zero-range force
($\lambda=\infty$). In this instance, always with $k_F=1.71$~fm$^{-1}$, which
corresponds to a neutron density $\rho=0.17$~fm$^{-3}$, one gets
$V_{1,\mathrm{crit}}^{\mathrm{ring}}\cong-239$~MeV~fm$^3$ and
$V_{1,\mathrm{crit}}^{\mathrm{RPA}}\cong-159$~MeV~fm$^3$, 
respectively.
We remark that the value of $V_1^{\mathrm{ring}}$ for a zero range force
coincides with the one of Ref.~\cite{Hua98}, which in turn is identical to what
one gets in the Hartree theory, as we shall later see.

The values of the couplings in Eq.~(\ref{eq:V1crit}) signal the occurrence of a
spontaneous symmetry breaking: correspondingly, a phase transition takes place
in the system from a phase fully symmetric to a phase where the rotational
invariance is broken in spin space.
Hence, the vacuum appropriate to this new phase should be characterized by two
Fermi momenta $k_F^+$ and $k_F^-$, associated with neutrons with spin up and
down, respectively.

In this situation the free Fermi propagator, while still diagonal in spin
space, is no longer proportional to the unit matrix, but becomes 
\begin{equation}
  G^{0,\mathrm{b}}(\vec{k},\omega) = \left( 
    \begin{array}{cc}
      G^{0,\mathrm{b}}_{++}(\vec{k},\omega) & 0 \\
      0 & G^{0,\mathrm{b}}_{--}(\vec{k},\omega)
    \end{array} \right),
\end{equation}
where
\begin{subequations}
\label{eq:G0+-}
\begin{eqnarray}
  G^{0,\mathrm{b}}_{++}(\vec{k},\omega) &=& 
    \frac{\theta(k-k_F^+)}{\omega-\omega_{\vec{k}}+i\eta} + 
    \frac{\theta(k_F^+-k)}{\omega-\omega_{\vec{k}}-i\eta} , \\
   G^{0,\mathrm{b}}_{--}(\vec{k},\omega) &=& 
    \frac{\theta(k-k_F^-)}{\omega-\omega_{\vec{k}}+i\eta} + 
    \frac{\theta(k_F^--k)}{\omega-\omega_{\vec{k}}-i\eta}
\end{eqnarray}
\end{subequations}
and $\omega_{\vec{k}}=\hbar k^2/2m=\epsilon^{(0)}_{\vec{k}}$.

With the above propagator the first order momentum independent self-energy
is easily computed and, as expected, splits into two terms, namely
\begin{subequations}
\label{eq:Sigma+-}
\begin{eqnarray}
  \hbar\Sigma_{++} &=& \frac{V_1}{6\pi^2}\left[\left({k_F^+}^3-{k_F^-}^3\right)
    - \left({k_F^+}^3+2{k_F^-}^3\right)\right] \\
  \hbar\Sigma_{--} &=& \frac{V_1}{6\pi^2}\left[\left({k_F^-}^3-{k_F^+}^3\right)
    - \left({k_F^-}^3+2{k_F^+}^3\right)\right] 
\end{eqnarray}
\end{subequations}
of obvious physical meaning.

In Eq.~(\ref{eq:Sigma+-}) both the direct (first term on the right hand side)
and the exchange (second term) contributions are neatly separated out. 
Note that the direct term is the same in both $\Sigma_{++}$ and $\Sigma_{--}$,
but for the sign, and moreover it vanishes, as it should, for $k_F^+=k_F^-$.

Since for a homogeneous infinite system the HF problem is trivial at $T=0$,
{\em even in a broken vacuum}, then the exact HF fermion propagator is easily
obtained by replacing Eq.~(\ref{eq:G0+-}) with
\begin{subequations}
\label{eq:GHF+-}
\begin{eqnarray}
  \hskip -0.8cm G^{\mathrm{HF},\mathrm{b}}_{++}(\vec{k},\omega) &=& 
    \frac{\theta(k-k_F^+)}{\omega-(\omega_{\vec{k}}-3V_1{k_F^-}^3/6\pi^2\hbar)
    +i\eta} + 
    \frac{\theta(k_F^+-k)}{\omega-(\omega_{\vec{k}}-3V_1{k_F^-}^3/6\pi^2\hbar)
    -i\eta} , \nonumber \\ \\
  \hskip -0.8cm G^{\mathrm{HF},\mathrm{b}}_{--}(\vec{k},\omega) &=& 
    \frac{\theta(k-k_F^-)}{\omega-(\omega_{\vec{k}}-3V_1{k_F^+}^3/6\pi^2\hbar)
    +i\eta} + 
    \frac{\theta(k_F^--k)}{\omega-(\omega_{\vec{k}}-3V_1{k_F^+}^3/6\pi^2\hbar)
    -i\eta}. \nonumber \\
\end{eqnarray}
\end{subequations}
The above, to be viewed as a generalized HF fermion propagator in 
a vacuum not invariant for spin rotations, embodies the HF single-particle
energies 
\begin{equation}
\label{insiemepm}
  \epsilon_{\vec{k}}^{\pm} = \epsilon_{\vec{k}}^{(0)} -
    \frac{3V_1}{6\pi^2}{k_F^{\mp}}^3 = \epsilon_{\vec{k}}^{(0)} - \frac{3}{2}
    V_1 \rho(1\mp M)
\end{equation}
and allows to compute the HF energy per particle in terms of the relative
magnetization $M=(N^\uparrow-N^\downarrow)/(N^\uparrow+N^\downarrow)$ and of
the density $\rho=(N^\uparrow+N^\downarrow)/\Omega$ according to 
\begin{eqnarray}
\label{eq:E/N}
  \frac{E}{N} &=& \frac{1}{\rho} \int \frac{{\d}\vec{k}}{(2\pi)^3}
    \left[ \epsilon_{\vec{k}}^{(0)} - \frac{3}{2} V_1 \frac{{k_F^-}3}{6\pi^2}
    \right] \theta(k_F^+-k) \nonumber \\ && \quad + \frac{1}{\rho} \int
    \frac{{\d}\vec{k}}{(2\pi)^3} 
    \left[ \epsilon_{\vec{k}}^{(0)} - \frac{3}{2} V_1 \frac{{k_F^+}3}{6\pi^2}
    \right] \theta(k_F^--k) \nonumber \\
  &=& \frac{3}{10} \epsilon_F^{(0)} \left[ (1+M)^{5/3} + (1-M)^{5/3} \right]
    - \frac{3}{4} V_1 \rho (1-M^2) .
\end{eqnarray}
Here $N^\uparrow/\Omega={k_F^+}^3/6\pi^2$ and
$N^\downarrow/\Omega={k_F^-}^3/6\pi^2$ correspond to the spin up and spin down
densities, respectively.
Likewise, one can easily write down the expressions for
$\epsilon^{\pm}_{\vec{k}}$ and $E/N$ in Hartree and in Fock approximation
separately. We remark that Eq.~(\ref{insiemepm}) coincides with the
single-particle energies obtained by solving Eq.~(\ref{sistematrino}), with a 
zero-range force, in the limit $T\to0$, $B\to0$.

Now, for the system to be in equilibrium the two chemical potentials
(Fermi energies at $T=0$) of the spin up and spin down fermions should be
equal. Hence, in HF the equation
\begin{equation}
  \frac{\hbar^2{k_F^+}^2}{2m} - \frac{3 V_1}{6\pi^2} {k_F^-}^3 =
    \frac{\hbar^2{k_F^-}^2}{2m} - \frac{3 V_1}{6\pi^2} {k_F^+}^3 
\end{equation}
should be fulfilled and similar equations should hold in the Hartree and in the
Fock approximation as well. Notably all the three cases can be compactified
into the single equation 
\begin{equation}
\label{eq:Stoner}
  (1+M)^{2/3} - (1-M)^{2/3} = - \alpha \frac{V_1 \rho}{\epsilon_F{(0)}} M,
\end{equation}
with
\begin{equation}
\label{eq:alphaSt}
  \begin{array}{ll}
    \alpha=3 & \mathrm{in~ HF} \\
    \alpha=2 & \mathrm{in~ Hartree} \\
    \alpha=1 & \mathrm{in~ Fock} . 
  \end{array}
\end{equation}
In the context of solid state physics Eq.~(\ref{eq:Stoner}) is usually
referred to as {\em Stoner equation}.
Of course, the values in Eq.~(\ref{eq:alphaSt}) for $\alpha$ are valid only for
a zero-range force. 

It is a remarkable occurrence that one arrives exactly at the same
Eq.~(\ref{eq:Stoner}) by minimizing the energy per particle in
Eq.~(\ref{eq:E/N}) with respect to the magnetization. This outcome is a
consequence of the Hugenholtz-Van~Hove theorem \cite{Hug58}, which remains true
even in a broken vacuum. 

Now, the Stoner equation in Eq.~(\ref{eq:Stoner}) admits real solutions only
for $M$ in the range $0\le M\le1$. For a given density $\rho$, which also fixes
$\epsilon_F^{(0)}$, the case $M=1$, corresponding to a fully magnetized
system, leads to the critical value for the coupling
\begin{equation}
\label{eq:V1upper}
  V_{1,\mathrm{crit}}^{\mathrm{upper}} = - \frac{2^{2/3}}{\alpha}
    \frac{\epsilon_F^{(0)}}{\rho}.
\end{equation}
This expression {\em cannot} be derived in the linear response framework
and with $\alpha=2$ (Hartree approximation) is identical to the
mean field value quoted in Ref.~\cite{Hua98}. 

Instead, in the case $M=0^+$ (incipient ferromagnetism) Eq.~(\ref{eq:Stoner})
leads to 
\begin{equation}
\label{eq:V1lower}
  V_{1,\mathrm{crit}}^{\mathrm{lower}} = - \frac{4}{3\alpha}
    \frac{\epsilon_F^{(0)}}{\rho}. 
\end{equation}
This result, when $\alpha=2$ (Hartree approximation), coincides with the linear
response value in ring approximation in Eq.~(\ref{eq:V1critring}), whereas,
when $\alpha=3$ (HF approximation), coincides with the linear response value in
RPA in Eq.~(\ref{eq:V1critRPA}). 

It is clear that with the propagator in Eq.~(\ref{eq:GHF+-}) it would now be
possible to study the system's response to an external probe when the vacuum is
broken and the associated collective modes. These split into excitations along
the $z$-axis (assumed as the direction along which the spontaneous symmetry
breaking has taken place) and along the directions orthogonal to it (Goldstone
modes). This topic is presently addressed in a work in progress.

We anticipate, however, that we expect to get in this case a collective mode
characterized by a \textit{quadratic} --- not \textit{linear} as in
Eq.~(\ref{eq:magnons}) --- dispersion relation, which is the one actually
exhibited by ferromagnetic crystals. Our guess is supported by the findings of
Ref.~\cite{Alb89}, where, for a system (asymmetric nuclear matter) displaying a
spontaneous symmetry breaking in the isospin space, a collective mode with a
quadratic dispersion relation has indeed been found.

\section{Conclusions}

In this work we have reexamined the issue of the connection between the
Matsubara theory at finite $T$ and the Feynman one at $T=0$ addressing a
specific example, namely an infinite homogeneous system of nucleons, and a
specific theoretical many-body framework, namely the HF mean field.

Our focus has been on how the theorem of Kohn, Luttinger and Ward is
actually implemented. In fact, the KLW theorem assures the identity between the
$T=0$ Feynman theory and the $T\to0$ limit of the Matsubara one (at least
for systems made up with spin one-half fermions) when no symmetry breaking
occurs. 

Yet, one would like to know the actual temperature behavior of the HF
diagrams of order higher than one: these in fact rigorously vanish at $T=0$ ---
for a static interaction ---
when computed as Feynman diagrams, but not when computed as $T\to0$ Matsubara
diagrams. Thus the theorem is realized through a cancellation, which becomes
``complete'' at $T=0$ and ``minimal'' at the Fermi temperature $T_F$.
Hence, here is where the HF problem for an infinite system becomes nearly as 
complex as in finite systems: indeed, reaching self-consistency at $T_F$
requires to account for many diagrams beyond the first order ones or,
equivalently, to perform many iterations when searching for a solution of the
non-linear, non-local HF equations.
Just how many it depends, of course, on the density and on the interaction and,
concerning the latter, more on its range than on its strength: in fact, for a
contact force the HF problem in an infinite system at any $T$ remains as
trivial as it is at $T=0$, namely it simply amounts to first order perturbation theory.

It becomes trivial as well at large $T$, at least when the interaction is of
pure exchange (and hence of pure quantum) nature, which is the one we confine
ourselves to consider, for sake of illustration, in this paper: large $T$
indeed means classical physics.  In addition, the impact of any interaction is
no longer felt at large $T$, where the kinetic energy dominates.

When a symmetry is broken, so it is the KLW theorem. In the second part of
this work we have addressed this issue again, aiming at exploring more closely
how this occurrence is emerging in a specific case, meant to be illustrative of
the phenomenon in general. 
We have thus studied a system of neutrons placed in an external magnetic field
$B$, which dynamically breaks the rotational invariance of the system's
Hamiltonian in spin space. For the purpose of clarifying the physics we choose
the strength of $B$ very large.

Three basic features emerge from our analysis.
First, although our two-body potential is of purely exchange nature --- thus
entailing only a Fock term in the interaction matrix elements --- the presence
of $B$ restores the presence of a Hartree term as well, whose relevance
dependes upon the observable one is dealing with.
Thus, in our example, the role of the induced Hartree term is minor as far as
the chemical potential in concerned, but substantial in the magnetization.

Secondly, as $T\to0$ the successive iterations leading to the self-consistent
field no longer cancel each other --- as they would were the KLW theorem valid
--- rather they yield a net contribution, actually as large as it is at $T_F$,
which stays pretty constant in the temperature range from $T_F$ to $T=0$.
We have verified the above outcome for the case of the chemical potential in
the presence of an unrealistically large magnetic field; different values of
$B$ would affect the size of the contribution, but not its existence.

Finally, it turns out that the approach to the self-consistent solution is much
dependent upon the \textit{sign} of the exchange interaction: indeed, we have
found that a ferromagnetic coupling among the neutrons swiftly and smoothly
leads to the HF mean field, whereas an antiferromagnetic coupling not only
requires more iterations before reaching self-consistency, but also entails an
approach to the latter which ``oscillates'' around the self-consistent
solution. 

Given that in the presence of a dynamical symmetry breaking the Feynman and
Matsubara theories in the $T\to0$ limit provide different results and that
those of Matsubara are the correct ones, the question naturally arises: is it
possible to obtain the right expectation values for the observables also in the
framework of the zero-temperature theory?
The answer to this question is positive, provided the appropriate propagator is
used at $T=0$ to compute the observables. 

Thus, in the example we have treated, the chemical potential and the
magnetization obtained from the HF single-particle thermal propagator in a
magnetic field $\mathcal{G}_{\text{HF}}^B$ can be obtained as well in the
framework of Feynman theory at $T=0$ using the polarization propagator
$\Pi(\vec{q},\omega)$. The latter should however embody the resummation of an
infinite number of loops: specifically, the results obtained from
$\mathcal{G}_{\text{HF}}^B$ in the $T\to0$ limit are recoverd in the $T=0$
framework using $\Pi^{\text{RPA}}$. Actually this correspondence applies to any
infinite subset of many-body diagrams, in the sense, for instance, that
$\Pi^{\text{ring}}$ corresponds to $\mathcal{G}_{\text{H}}^B$ and
$\Pi^{\text{ladder}}$ (the steps of the ladder being between a particle and a
hole) corresponds to $\mathcal{G}_{\text{F}}^B$.
Of course, in the absence of any interaction the correspondence is between
$\Pi^0$ and $\mathcal{G}_0^B$.

However, the above statements are valid only in a limited range of values of
$B$, namely as long as the linear response theory is tenable.
It is worth noticing that we can actually compute this range by comparing the
predictions (e.~g., on the magnetization) of the single-particle thermal
propagator in the $T\to0$ limit with those of the polarization propagator.
It turns out that the range of applicability of the linear response theory is
not much affected by the many-body scheme employed.

A remarkable feature of the linear response theory lies in its ability to
predict the onset of a phase transition in the system, in our case of the
ferromagnetic phase, as a function not of the temperature, but rather of the
strength of the coupling constant $V_1$ viewed as a control parameter entering
into the system's Hamiltonian.
In fact, expressing the magnetic susceptibility of the system via the real part
of $\Pi(\vec{q},\omega)$ and searching for the poles of the latter in the
variable $V_1$, one finds \cite{Bro72}, both in RPA and in ring approximation,
the critical values $V_{1,\text{crit}}^{\text{RPA}}$ and
$V_{1,\text{crit}}^{\text{ring}}$ where the phase transition starts to occur
\textit{if the coupling is ferromagnetic} (incipient ferromagnetism).

Indeed, as it is well known, the spontaneous symmetry breaking is heralded by a
dramatic increase of the isothermal susceptibility and by a spectacular
enhancement of the spin response function at wavevectors small enough to
encompass the long range spin order taking place in the system.
Note that $V_{1,\text{crit}}^{\text{RPA}}$ and
$V_{1,\text{crit}}^{\text{ring}}$ are substantially different for a contact
force (by a factor 2/3, $V_{1,\text{crit}}^{\text{ring}}$ being the larger
one, showing that the phase transition is made easier by antisymmetrization),
while they become closer and closer as the range of the interaction increases,
as it should be.

For an antiferromagnetic coupling the above does not occur because, as already
anticipated, the Fermi gas, even in the absence of the interaction, lives in
the antiferromagnetic phase. Hence, it should display Goldstone modes with a
linear relationship between frequency and wavevector, which indeed we have
found. 

Naturally, one would like as well to explore the Goldstone modes in the
ferromagnetic case: this we have not done, but we have paved the way to their
investigation using the technique of the anomalous propagator. 
In fact, in the ferromagnetic phase the system develops two Fermi momenta,
$k_F^+$ and $k_F^-$, and, as a consequence, the single-particle propagator,
while still diagonal in the spin indices, is no longer proportional to the unit
matrix, but splits into the two components $G_{++}^{\text{HF},b}$ and
$G_{--}^{\text{HF},b}$. These we have computed and, with their help, we have
computed as well the system's energy per particle in terms of the magnetization
$M$. By minimizing the energy with respect to $M$ or, equivalently, by
requiring identical chemical potential for the two spin species of particles,
we have found the critical value of the coupling constant,
$V_{1,\text{crit}}^{\text{upper}}$, yielding a fully magnetized system, a
result impossible to reach in the standard linear response theory framework.
Worth mentioning is that the HF value of $V_{1,\text{crit}}^{\text{upper}}$ is
lower than the corresponding values obtained in the simpler schemes of the
Hartree's or Fock's theories.

The above results clearly would not be attainable in the scheme of the Feynman
perturbative theory at $T=0$, even using the polarization propagator, which of 
course illustrates that in the presence of a spontaneous symmetry breaking the
KLW theorem does not hold.

In conclusion, it appears worth remarking that the \textit{free} Fermi gas, a
system to be viewed at $T=0$ as antiferromagnetic owing to its spin-zero wave
function, when the interaction is switched on displays both in the
ferromagnetic and in the antiferromagnetic phases the same magnons found in
crystals, in spite of being a continuous system.
This recognition offers the opportunity to establish a contact between the
computations performed on the lattice and within advanced many-body frameworks
based on the Fermi gas model, for example in connection with the Heisenberg
Hamiltonian. Not however with the Ising model, which, although dealing with a 
physics not so far from the one dealt with in this paper, actually
does not display the variety of collective (Goldstone) modes showing up in the
present system. This occurrence reflects the fact that what is broken is a
discrete symmetry in the Ising model, a continuous one in our infinite,
homogeneous system of neutrons. 

\appendix

\section{}
\label{app:mu3}

In this Appendix we recall the power expansion for the chemical potential of a
free Fermi gas as a function of $k_BT/\epsilon_F$ to the order 
$O(k_BT/\epsilon_F)^7$.

Starting from the Sommerfeld's expansion of the density \cite{Pat72},
\begin{eqnarray}
\label{appdensita`}
  \rho&=&\frac{(2m\epsilon_{F})^{3/2}}{3\pi^2\hbar^3} \\ \nonumber 
  &=&\frac{(2m\mu)^{3/2}}{3\pi^2\hbar^3}
    \left[1+\frac{\pi^2}{8}\left(\frac{k_BT}{\mu}\right)^2+
    \frac{7\pi^4}{640}\left(\frac{k_BT}{\mu}\right)^4+
    \frac{31\pi^6}{3072}\left(\frac{k_BT}{\mu}\right)^6+... \right],
\end{eqnarray}
one solves the equation 
\begin{equation}
\label{appmuimplicito}
  \mu=\epsilon_{F}\left[1+\frac{\pi^2}{8}\left(\frac{k_BT}{\mu}\right)^2+
    \frac{7\pi^4}{640}\left(\frac{k_BT}{\mu}\right)^4+
    \frac{31\pi^6}{3072}\left(\frac{k_BT}{\mu}\right)^6 + ... \right]^{-2/3}
\end{equation}
iteratively. At third order in $T^2$ (the zero order being $\mu=\epsilon_F$)
one obtains 
\begin{equation}
\label{appmusviluppo}
  \mu=\epsilon_{F}\left[1-\frac{\pi^2}{2^2~3}
    \left(\frac{k_BT}{\epsilon_{F}}\right)^2- 
    \frac{\pi^4}{2^4~5}\left(\frac{k_BT}{\epsilon_{F}}\right)^4-
    \frac{247~\pi^6}{2^6~3^4~5}\left(\frac{k_BT}{\epsilon_{F}}\right)^6+...
    \right],
\end{equation}
the last term on the right hand side not being quoted in the literature to our
knowledge. 

\section{}
\label{app:exactRPA}

As shown in Ref.~\cite{Alb82a}, the RPA polarization propagator ---
approximately expressed by Eq.~(\ref{eq:PiRPA}) in the first order of the
continued fraction expansion --- is exactly given in the long wavelength, low
frequency limit by the formula
\begin{equation}
\label{eq:Pix}
  \lim_{x\to0} \Pi(x) = -\frac{m^*k_F}{(2\pi\hbar)^2}
    \frac{4}{1+\widetilde{u}_0}, 
\end{equation}
with $x=(\omega/q)m^*/\hbar k_F$. In Eq.~(\ref{eq:Pix}), $\widetilde{u}_0$ is
the amplitude of the $l=0$ partial wave entering into the expansion of the ph
matrix element
\begin{eqnarray}
\label{uph}
  \mathcal{U}_{\text{ph}}(\xi,\xi') &=& \frac{1}{2\pi} \int
    {\d}(\varphi_{\vec{k}}-\varphi_{\vec{k}'})
    V_{\text{ph}}^{\text{exch}}(\vec{k},\vec{k}')_{k=k'=k_F} \nonumber \\
  &=& \frac{(2\pi\hbar)^2}{2m^{\star}k_F}\sum_{l=0}^{M}(2l+1) 
    \widetilde{u}_{l}P_l(\xi)P_l(\xi'),
\end{eqnarray}
which is valid close to the Fermi surface (in general
$\mathcal{U}_{\text{ph}}(\xi,\xi')$ may have an arbitrarily large number $M$ of
components). 

In Eq.~(\ref{uph}) $\xi=\widehat{k}\cdot\widehat{q}$ and
$\xi'=\widehat{k'}\cdot\widehat{q}$, $\widehat{k}$ and $\widehat{k'}$ being the
directions of the momenta of the initial and final hole, respectively, and
$\widehat{q}$ the direction of the momentum transfer. Moreover, in
Eq.~(\ref{uph}) the integration is over the azimuthal angles of $\vec{k}$ and
$\vec{k'}$ and the $P_l$ are the Legendre polynomials. 

Now, with the simple interaction we use, the $\sigma=1$ (spin one) ph matrix
element reads
\begin{equation}
  V_{\text{ph}}(\vec{k},\vec{k}',\vec{q}) = 
    2 V_1 \frac{\lambda}{\lambda^2+q^2} + 
    V_1\frac{\lambda^2}{\lambda^2+|\vec{k}-\vec{k}'|^2},  
\end{equation}
where the direct and the exchange contributions are explicitly displayed. 
The direct one, easy to deal with, leads to the ring approximation expression
in Eq.~(\ref{chiring}). Concerning the exchange term, from Eq.~(\ref{uph}) one
finds 
\begin{equation}
\label{uint}
  \mathcal{U}_{\text{ph}}^{\text{ex}}(\xi,\xi') = V_1 \left(1+4 
    \frac{(\xi-\xi')^2}{\bar\lambda^4}+4 \frac{1-\xi\xi'}{\bar\lambda^2}
    \right)^{-1/2} ,
\end{equation}
again with $\bar\lambda=\lambda/k_F$.

By comparing then Eqs.~(\ref{uph}) and (\ref{uint}) one finally gets 
\begin{eqnarray}
  \widetilde{u}_{0} &=& \frac{m^*k_F}{(2\pi\hbar)^2} \frac{V_1}{2} 
    \int_{-1}^{1}\d\xi  \int_{-1}^{1}\d\xi' \left(1+4 
    \frac{(\xi-\xi')^2}{\bar\lambda^4}+4 \frac{1-\xi\xi'}{\bar\lambda^2}
    \right)^{-1/2} \nonumber \\
  &=& \frac{3\rho}{2\epsilon_F^{\star}} 
    \frac{V_1}{2}\left(\frac{\bar\lambda}{2}\right)^2 
    \ln\left[1+\left(\frac{2}{\bar\lambda}\right)^2\right]
\end{eqnarray}
($\rho=k_F^3/3\pi^2$ being the density of our homogeneous neutron system), from
which Eq.~(\ref{eq:chiRPA}) is obtained.

\end{document}